\documentclass{article}%
\usepackage{graphicx}
\usepackage{amsmath}%
\usepackage{amsfonts}%
\usepackage{amssymb}
\renewcommand{\mathbf}[1]{\mbox{\boldmath$#1$}}
\newtheorem{theorem}{Theorem}
\newtheorem{acknowledgement}[theorem]{Acknowledgement}

\begin{document}

\title{{\Large Symmetries and physical functions in general gauge theory}}
\author{D.M. Gitman\thanks{Institute of Physics, University of S\~{a}o Paulo, Brazil;
e-mail: gitman@dfn.if.usp.br} \ and I.V. Tyutin\thanks{Lebedev Physics
Institute, Moscow, Russia; e-mail: tyutin@lpi.ru}}
\date{\today              }
\maketitle

\begin{abstract}
The aim of the present article is to describe the symmetry structure of a
general gauge (singular) theory, and, in particular, to relate the structure
of gauge transformations with the constraint structure of a theory in the
Hamiltonian formulation. We demonstrate that the symmetry structure of a
theory action can be completely revealed by solving the so-called symmetry
equation. We develop a corresponding constructive procedure of solving the
symmetry equation with the help of a special orthogonal basis for the
constraints. Thus, we succeed in describing all the gauge transformations of a
given action. We find the gauge charge as a decomposition in the orthogonal
constraint basis. Thus, we establish a relation between the constraint
structure of a theory and the structure of its gauge transformations. In
particular, we demonstrate that, in the general case, the gauge charge cannot
be constructed with the help of some complete set of first-class constraints
alone, because the charge decomposition also contains second-class
constraints. The above-mentioned procedure of solving the symmetry equation
allows us to describe the structure of an arbitrary symmetry for a general
singular action. Finally, using the revealed structure of an arbitrary gauge
symmetry, we give a rigorous proof of the equivalence of two definitions of
physicality condition in gauge theories: one of them states that physical
functions are gauge-invariant on the extremals, and the other requires that
physical functions commute with FCC (the Dirac conjecture).
\end{abstract}

\section{Introduction}

Most of the contemporary particle-physics theories are formulated as gauge
theories. It is well-known that within the Hamiltonian formulation gauge
theories are theories with constraints, in particular, first-class constraints
(FCC). This fact is the main reason for a long and intensive study of the
formal theory of constrained systems. The theory of constrained systems,
initiated by the pioneering works of Bergmann and Dirac \cite{Berg,Dirac} and
then developed and presented in various review books
\cite{Sudar74,HanReT76,Sunde82,GitTy90,HenTe92}, still attracts a great
attention of researchers. Relatively simple were the first steps of the theory
in formulating the dynamics of constrained systems in the phase space, thus
elaborating the procedure of finding all the constraints (Dirac's procedure)
and reorganizing the constraints to FCC and second-class constraints (SCC).
From the very beginning, it became clear that the presence of FCC among the
complete set of constraints in the Hamiltonian formulation is a direct
indication that the theory is a gauge one, i.e., its Lagrangian action is
invariant under gauge transformations, which, in the general case, are
continuos transformations parameterized by arbitrary functions of time (of the
space-time coordinates, in the case of field theory). It was demonstrated that
the number of independent gauge parameters is equal to the number $\mu_{1}$ of
primary FCC, and the total number of nonphysical variables is equal to the
number $\mu$ of all FCC, despite the fact that the equations of motion contain
only $\mu_{1}$ arbitrary functions of time (undetermined Lagrange multipliers
to the primary FCC); see \cite{GitTy90} and references therein. At the same
time, we proved that for a class of theories for which the constraint
structures of the complete theory and of its quadratic approximation are the
same, and for which the constraint structure does not change from point to
point in the phase-space (we call such theories perturbative ones), physical
functions in the Hamiltonian formulation must commute with FCC. In a sense,
this statement can be identified with the so-called Dirac conjecture. All
models known until now in which the Dirac conjecture does not hold are
nonperturbative in the above sense. After this preliminary progress in the
theory of constrained systems, it became clear that a natural and very
important continuation of studies is the attempt to relate the constraint
structure and constraint dynamics of a gauge theory in the Hamiltonian
formulation to specific features of a theory in the Lagrangian formulation,
especially to relate the constraint structure to the structure of gauge
transformations for a Lagrangian action. One of the key problems here is the
following: how to construct a general expression for the gauge charge if the
constraint structure in the Hamiltonian formulation is known? Another
principle question closely related to the latter is: can one identify the
physical functions defined as those commuting with FCC in the Hamiltonian
formulation (the so-called physicality condition in the Hamiltonian sense,
well-known as the Dirac conjecture) with the physical functions defined as
gauge-invariant functions? (In what follows, we refer to the latter definition
of physical functions as the physicality condition in the gauge sense.) Many
efforts have been made in attempting to answer these questions: see, for
example, \cite{Castell}. All previous considerations contain some restricting
assumptions about the structure of a theory (in particular, about the
constraint structure), so that rigorous answers to all of the above questions
are still unknown for a general gauge theory (even one belonging to the
above-mentioned perturbative class). The aim of the present work is to answer
the above questions for perturbative gauge theories in terms of rigorous statements.

In this connection, it should be noted that there exists an isomorphism
between the symmetry classes of the Hamiltonian and Lagrangian actions of the
same theory, since they are dynamically equivalent\footnote{Suppose that an
action $S[q,y]$ contains two groups of coordinates $q$ and $y$, such that the
coordinates $y$ can be expressed as local functions $y=\bar{y}\left(
q^{\left[  l\right]  },l<\infty\right)  \,$of $q$ and their time derivatives
with the help of the equations $\delta S/\delta y=0$. We call $y$ auxiliary
coordinates. The action $S[q,y]$ and the reduced action $S\left[  q\right]
=S[q,\bar{y}]$ lead to the same equations for the coordinates $q$, see
\cite{Superflu,BorTy98}. The actions $S[q,y]$ and $S\left[  q\right]  $ are
called dynamically equivalent actions.} (see \cite{GitTy05b}). It is then
convenient to study the symmetry structure by considering the simpler
Hamiltonian action (this is what we do in the present article), since it is a
first-order action. The symmetries of the Lagrangian action can be obtained as
a reduction of the Hamiltonian action symmetries by substituting all the
Lagrange multipliers and momenta for coordinates and velocities. We
demonstrate that the symmetry structure of the Hamiltonian action (and, hence,
that of the Lagrangian action) can be completely revealed by solving the
so-called symmetry equation. Choosing a special orthogonal basis for the
constraints (introduced in \cite{196} and described in sec. 3), one can
analyze the symmetry equation algebraically. We develop the corresponding
constructive procedure of solving the symmetry equation. Thus, we succeed in
describing all the gauge transformations of a given action (sec. 4). We find
the gauge charge as a decomposition in the orthogonal constraint basis. Thus,
we establish the relation between the constraint structure of a theory and the
structure of its gauge transformations. In particular, we demonstrate that, in
the general case, the gauge charge cannot be constructed with the help of some
complete set of FCC alone, since the decomposition also contains SCC. The
above-mentioned procedure of solving the symmetry equation allows one to
analyze the structure of any infinitesimal Noether symmetry. In doing this, we
constructively demonstrate that any infinitesimal Noether symmetry can be
represented as a sum of three kinds of symmetries: global, gauge, and trivial
(sec. 5). In particular, we can see that the global part of a symmetry is a
canonical transformation, which does not vanish on the extremals, and the
corresponding conserved charge (the generator of this transformation) does not
vanish on the extremals either. The gauge part of a symmetry does not vanish
on the extremals, but the gauge charge does vanish on them. The trivial part
of any symmetry vanishes on the extremals, and the corresponding charge
vanishes quadratically on the extremals. In our procedure of solving the
symmetry equation, the generators of canonical global and gauge symmetries may
depend on Lagrange multipliers and their time derivatives. This happens in the
case when the number of stages in the Dirac procedure is more than two. In
addition, we prove that any infinitesimal Noether symmetry that vanishes on
the extremals is a trivial symmetry (sec. 6). Finally, using the established
structure of an arbitrary gauge symmetry, we strictly prove an equivalence of
the two definitions of physicality condition in gauge theories: one of them
states that physical functions are gauge-invariant on the extremals, and the
other requires that physical functions commute with FCC (the Dirac conjecture)
(sec. 7). In sec. 2, we present the basic notation and definitions.

\section{Basic notation and definitions}

We consider finite-dimensional systems described by a set of generalized
coordinates\emph{ }$q\equiv\{q^{a};\;a=1,2,...,n\}$. The following notation is
used:
\begin{equation}
q^{a\left[  l\right]  }=\left(  d_{t}\right)  ^{l}q^{a}%
\,,\;\;l=0,1,...,\;\;\left(  q^{a\left[  0\right]  }=q^{a}\right)
\,,\;\;d_{t}=\frac{d}{dt}\,. \label{1.1}%
\end{equation}
We recall that the space of $q^{a\left[  l\right]  }$, $a=1,...,n$%
,$\;l=0,1,...,N_{a}\,$,\ regarded as independent variables, is called the jet
space \cite{Kuper79}. The majority of physical quantities in classical
mechanics are described by so-called local functions (LF) which are defined on
the jet space. LF depend on $q^{a\left[  l\right]  }$,$\;l\leq N_{a}$%
,$\;$where $N_{a}<\infty$\thinspace.\ We often denote LF as\footnote{Functions
$F$ may depend on time explicitly; however, we do not include $t$ in the
arguments.}
\begin{equation}
F\left(  q^{a\left[  0\right]  },q^{a\left[  1\right]  },q^{a\left[  2\right]
},...\right)  =F\left(  q^{[]}\right)  \,. \label{1.2}%
\end{equation}

In our considerations, we use so-called local operators (LO). An LO
$\hat{U}_{Aa}$ is a matrix operator which acts on columns of LF $f^{a}$
producing columns $F_{A}$ of LF, $F_{A}=\hat{U}_{Aa}f^{a}\,$. Such LO have the
form
\begin{equation}
\hat{U}_{Aa}=\sum_{k=0}^{K<\infty}u_{Aa}^{k}\left(  d_{t}\right)  ^{k}\,,
\label{1.3}%
\end{equation}
where $u_{Aa}^{k}=u_{Aa}^{k}\left(  q^{[]}\right)  $ are LF. The operator
\begin{equation}
\left(  \hat{U}^{T}\right)  _{aA}=\sum_{k=0}^{K<\infty}\left(  -d_{t}\right)
^{k}u_{Aa}^{k} \label{1.4}%
\end{equation}
is called the operator transposed to $\hat{U}_{Aa}$. The relation%
\begin{equation}
F^{A}\hat{U}_{Aa}f^{a}=\left[  \left(  \hat{U}^{T}\right)  _{aA}F^{A}\right]
f^{a}+d_{t}Q\,, \label{1.5}%
\end{equation}
where $Q$ is an LF, holds for any LF $F^{A}$ and $f^{a}$. The LO
$\hat{U}_{ab}$ is symmetric ($+$) or antisymmetric ($-$), respectively, if
$\left(  \hat{U}^{T}\right)  _{ab}=\pm\hat{U}_{ab}\,.$ Thus, for any
antisymmetric LO $\hat{U}_{ab}$, relation (\ref{1.5}) implies $f^{a}%
\hat{U}_{ab}f^{b}=dQ/dt\,$.

We say that $F_{A}\left(  q^{[]}\right)  =0$ and $\chi_{\alpha}\left(
q^{[]}\right)  =0$ are equivalent sets of equations (and denote this fact as
$F=0\Longleftrightarrow\chi=0$) whenever they have the same sets of solutions.
By $O\left(  F\right)  $ we denote any LF that vanishes on the equations
$F_{a}\left(  q^{[]}\right)  =0$. More exactly, we define $O\left(  F\right)
=\hat{V}^{a}F_{a}\,,$ where $\hat{V}^{b}$ are LO. Besides, we denote via
$\hat{U}=\hat{O}\left(  F\right)  $ any LO that vanishes on the equations
$F_{a}\left(  q^{[]}\right)  =0$. That means that LF $u$ that enter into the
representation (\ref{1.3}) for such an operator vanish on these equations,
$u=O\left(  F\right)  $, or, equivalently $\hat{U}f=O\left(  F\right)  $ for
any LF $f.$

We consider Lagrangian theories given by an Lagrangian action $S\left[
q\right]  ,$
\begin{equation}
S\left[  q\right]  =\int_{t_{1}}^{t_{2}}Ldt\,,\;L=L\left(  q^{[]}\right)  \,,
\label{1.6}%
\end{equation}
where the Lagrange function $L$ is defined as an LF on the jet space. The
Euler--Lagrange equations are
\begin{equation}
\frac{\delta S}{\delta q^{a}}=\sum_{l=0}\left(  -d_{t}\right)  ^{l}%
\frac{\partial L}{\partial q^{a\left[  l\right]  }}=0\,. \label{1.7}%
\end{equation}

Any LF of the form $O\left(  \delta S/\delta q\right)  $ is called an extremal.

An infinitesimal transformation $q\rightarrow q^{\prime}=q+\delta q,$ with
$\delta q=\delta q\left(  q^{[]}\right)  $, being an LF, is a symmetry of $S$
in case%
\begin{equation}
\delta L=d_{t}F\,, \label{1.9}%
\end{equation}
where $F\left(  q^{[]}\right)  $ is an LF (such transformations are called
Noether symmetries).

Any infinitesimal symmetry implies a conservation law\textbf{ (}Noether
theorem)\textbf{:}%
\begin{equation}
\delta q^{a}\frac{\delta S}{\delta q^{a}}+d_{t}G=0\,,\;G=P-F\,. \label{1.12}%
\end{equation}
Here, the LF $F$ and $P$ are
\begin{equation}
\delta L=d_{t}F\,,\;P=\sum\nolimits_{a}\sum_{m=1}^{N_{a}}p_{a}^{m}\delta
q^{a\left[  m-1\right]  }\,,\;p_{a}^{m}=\sum_{s=l}^{N_{a}}\left(
-d_{t}\right)  ^{^{s-m}}\frac{\partial L}{\partial q^{a\left[  s\right]  }}\,,
\label{1.13}%
\end{equation}
and the local function $G$ (which is constant on the extremals) is referred to
as a conserved charge\emph{\ }related to the symmetry $\delta q$. In what
follows, we call (\ref{1.12}) the symmetry equation. The quantities $\delta
q,\,S,$ and $G$ are related by the symmetry equation. Here and elsewhere, we
often call the set $\delta q$, $G$ a symmetry of the action $S.$

Noether infinitesimal symmetries can be global, gauge and trivial. Global
symmetries have nonzero (on the extremals) conserved charges; they are
parameterized by a set of time-independent parameters $\nu^{\alpha}$%
,\ $\alpha=1,...,r$, and have the form $\delta q^{a}(t)=\rho_{\alpha}%
^{a}(t)\nu^{\alpha},$ where $\rho_{\alpha}^{a}(t)$ are generators of the
global symmetry transformations. Gauge symmetries have zero (on the extremals)
conserved charges; they are parameterized by time-dependent gauge parameters
$\nu^{\alpha}\left(  t\right)  $, $\alpha=1,...,r$, which are arbitrary
functions of time (in the case of field theory, the gauge parameters depend on
all space-time variables). Infinitesimal gauge transformations have the form%
\begin{equation}
\delta q^{a}(t)=\hat{\Re}_{\alpha}^{a}\left(  t\right)  \nu^{\alpha}(t),
\label{1.10}%
\end{equation}
where the operators $\hat{\Re}_{\alpha}^{a}\left(  t\right)  $ are called
generators of gauge transformations.

For any action, there exist trivial symmetry transformations $\delta
_{\mathrm{tr}}q$,%
\begin{equation}
\delta_{\mathrm{tr}}q^{a}=\hat{U}^{ab}\frac{\delta S}{\delta q^{b}}\,,
\label{1.11}%
\end{equation}
where $\hat{U}$\ is an antisymmetric LO,$\left(  \hat{U}^{T}\right)
^{ab}=-\hat{U}^{ab}$.

Trivial symmetries have zero (on the extremals) conserved charges; see below.
Trivial symmetry transformations do not affect genuine trajectories. Two
symmetry transformations $\delta_{1}q$ and $\delta_{2}q$ are called equivalent
($\delta_{1}q\sim\delta_{2}q$)\ whenever they differ by a trivial symmetry
transformation: $\delta_{1}q\sim\delta_{2}q\Longleftrightarrow\delta
_{1}q-\delta_{2}q=\delta_{\mathrm{tr}}q\,.$ Thus, all the symmetry
transformations of an action $S$ can be divided into\emph{\ }equivalence classes.

\section{Symmetry equation and orthogonal constraint basis}

We consider here the Hamiltonian formulation of a singular theory
\cite{GitTy90,195}. The corresponding Hamiltonian action is denoted by
$S_{\mathrm{H}}$. We denote all irreducible constraints of the theory via
$\Phi=\left(  \Phi_{l}\left(  \eta\right)  \right)  $, where $\eta=\left(
x,p\right)  $ are all phase-space variables\footnote{In the general case of
theories with higher derivatives, the space of $x$ is larger than the space of
generalized coordinates $q$, see, e.g. \cite{195}.},%
\begin{align}
&  \eta^{A}=\left(  x^{a},p_{a}\right)  ,\;A=\left(  \alpha,a\right)
,\;\alpha=1,2,\;a=1,...,n\,,\nonumber\\
&  \mathrm{such\;that:\;\;}\eta^{1a}=x^{a}\,,\;\eta^{2a}=p_{a}\,.\,
\label{hs.5}%
\end{align}

We suppose that the constraints of the theory are reorganized so that they can
be divided into FCC, $\chi\left(  \eta\right)  $, and SCC, $\varphi\left(
\eta\right)  .$ Thus, $\Phi=\left(  \chi,\varphi\right)  .$ The Hamiltonian
action and the corresponding equations of motion are
\begin{align}
&  S_{\mathrm{H}}\left[  \mathbf{\eta}\right]  =\int\left[  p\dot
{x}-H^{\left(  1\right)  }\left(  \mathbf{\eta}\right)  \right]
dt\,,\;\mathbf{\eta}=\left(  \eta,\lambda\right)  \,,\nonumber\\
&  H^{\left(  1\right)  }\left(  \mathbf{\eta}\right)  =H\left(  \eta\right)
+\lambda\Phi^{\left(  1\right)  }\left(  \eta\right)  ;\nonumber\\
&  \frac{\delta S_{\mathrm{H}}}{\delta\mathbf{\eta}}=0\Longrightarrow\left\{
\begin{array}
[c]{l}%
\dot{\eta}=\{\eta,H^{\left(  1\right)  }\}\,\\
\Phi^{\left(  1\right)  }\left(  \eta\right)  =0\,
\end{array}
\right.  , \label{2.1}%
\end{align}
where $H^{(1)}$ is the total Hamiltonian, $\Phi^{\left(  1\right)  }$ are
primary\ constraints, and $\lambda$ are Lagrange multipliers to the primary constraints.

In the general case, the Hamiltonian $H$ and the constraints $\Phi$ can depend
on time $t$ explicitly. We take such a possibility into account. However, the
argument $t$ will not be written explicitly in what follows. Using Eq.
(\ref{2.1}), we can write the total time derivative of a function $f\left(
\eta\right)  $ as%
\[
d_{t}f=\partial_{t}f+\dot{\eta}\partial_{\eta}f=\left\{  f\,,H^{\left(
1\right)  }+\epsilon\right\}  +(\dot{\eta}-\{\eta,H^{\left(  1\right)
}\})\partial_{\eta}f\,.
\]

Here, $\epsilon$ is the (formal) momentum conjugate to time $t$, and the
Poisson brackets are defined in the extended phase space of the variables
$\eta;t,\epsilon$; for details, see \cite{GitTy90}.

If $\delta\mathbf{\eta=}\left(  \delta x,\delta p,\delta\lambda\right)  $ is a
symmetry of the Hamiltonian action $S_{\mathrm{H}}$, then the symmetry
equation is%
\begin{equation}
\delta\mathbf{\eta}\frac{\delta S_{\mathrm{H}}}{\delta\mathbf{\eta}}%
+d_{t}G=0\,, \label{2.2}%
\end{equation}
where $G$ is the corresponding conserved charge. One can study the symmetry of
the action $S_{\mathrm{H}}$ by solving this symmetry equation. If
$\delta\mathbf{\eta}$ is a symmetry of the Hamiltonian action $S_{\mathrm{H}}%
$, then symmetries $\delta q_{\mathrm{L}}\left(  q^{[]}\right)  $ of the
Lagrangian action $S$ can be obtained by writing the Hamiltonian symmetries
$\delta q\left(  \mathbf{\eta}^{[]}\right)  $, $\delta q\in\delta x$, as
functions on the jet space $q^{[]},$ see \cite{GitTy05b}.

For further consideration, especially for solving the symmetry equation, it is
convenient to accept that the set of all the constraints $\Phi$ is already
reorganized to the special form consistent with the Dirac procedure; see
\cite{196}. We call such a reorganized set of constraints an orthogonal
constraint basis. In this case,%
\begin{equation}
\Phi=\left(  \Phi^{\left(  i\right)  }\right)  ,\ \;\Phi^{\left(  i\right)
}=(\varphi^{(i)};\chi^{\left(  i\right)  })\,,\ i=1,...,\aleph\,. \label{2.3}%
\end{equation}
Such $\chi^{\left(  i\right)  }$ and $\varphi^{(i)}$ are, respectively, the
FCC and SCC of the $i$-th stage of the Dirac procedure, while $\aleph$ is the
number of the final stage. The total Hamiltonian is%
\begin{equation}
H^{\left(  1\right)  }=H+\lambda\Phi^{\left(  1\right)  }=H+\lambda_{\varphi
}\varphi^{\left(  1\right)  }+\lambda_{\chi}\chi^{\left(  1\right)  }\,,
\label{2.3a}%
\end{equation}
where $\chi^{\left(  1\right)  }$ are the primary FCC, and $\varphi^{\left(
1\right)  }$ are the primary SCC; $\lambda=\left(  \lambda_{\varphi}%
,\lambda_{\chi}\right)  $, while $\lambda_{\varphi}$ and $\lambda_{\chi}$ are
the corresponding Lagrange multipliers to the primary SCC and FCC,
respectively. At each stage of the Dirac procedure, the constraints are
divided into groups:
\begin{align}
\varphi^{\left(  i\right)  }  &  =\left(  \varphi^{\left(  i|s\right)
}\right)  \,,\;s=i,...,\aleph_{\varphi}\,;\nonumber\\
\chi^{\left(  i\right)  }  &  =\left(  \chi^{\left(  i|a\right)  }\right)
\,,\;a=i,...,\aleph_{\chi}\,, \label{2.4}%
\end{align}
where $\aleph_{\varphi}$ and $\aleph_{\chi}$ stand for the numbers of the
final stages at which SCC and new FCC, respectively, still appear. We can
write\footnote{The following notation is used: suppose that $F_{a}\left(
\eta\right)  ,\;a=1,...,n$, are some functions, then $\left[  F\right]  $ is
the number of these functions, $\left[  F\right]  =n$. Note that the brackets
$\left[  {}\right]  $ are also used to denote time derivatives ($q^{\left[
l\right]  }=\left(  d_{t}\right)  ^{l}q\,$) and the arguments of an action
functional (e.g. $S\left[  q\right]  $).}:%
\[
\lbrack\varphi^{(i)}]=0,\;i>\aleph_{\varphi}\,;\;\;[\chi^{\left(  i\right)
}]=0\,,\;i>\aleph_{\chi}\,,\;\aleph=\max\left(  \aleph_{\varphi}%
,\,\aleph_{\chi}\right)  \,.
\]
In what follows, we often use the notation%
\[
\lambda_{\varphi^{\left(  1|s\right)  }}=\lambda_{s}\,,\;\lambda
_{\chi^{\left(  1|a\right)  }}=\lambda^{a}\,,
\]
so that%
\begin{equation}
H^{\left(  1\right)  }=H+\lambda_{s}\varphi^{\left(  1\left|  s\right|
\right)  }+\lambda^{a}\chi^{\left(  1\left|  a\right|  \right)  }\,.
\label{2.5}%
\end{equation}
The division (\ref{2.4}) produces chains of constraints. All constraints in a
chain are of the same class. One ought to say that the numbers of constraints
at each stage in the same chain are equal. At the same time, each chain may
either be empty or contain several functions. Therefore, whenever there exist
FCC (SCC), the corresponding primary FCC (SCC) exist as well.

There exist $\aleph_{\varphi}$\ chains of SCC,%
\[
\varphi^{\left(  ...|s\right)  }=\left(  \varphi^{\left(  i|s\right)
}\,,\;i=1,...,s\right)  \,,\;s=1,...,\aleph_{\varphi}\,,
\]
labeled by the index $s,$ and $\aleph_{\chi}$\ chains of FCC,
\[
\chi^{\left(  ...|a\right)  }=\left(  \chi^{\left(  i|a\right)  }%
,\;i=1,...,a\right)  \,,\;a=1,...,\aleph_{\chi}\,,
\]
\textbf{\ }labeled by the index $a.$ The length of the longest chain of SCC is
$\aleph_{\varphi}$, and the length of the longest chain of FCC is
$\aleph_{\chi}$. The orthogonal constraint basis has the following properties:%

\begin{align}
&  \left.
\begin{array}
[c]{l}%
\left\{  \varphi^{\left(  i|s\right)  },\bar{H}+\epsilon\right\}
=\varphi^{\left(  i+1|s\right)  }+O\left(  \Phi^{\left(  ...i\right)  }\right)
\\
\left\{  \varphi^{\left(  i|s\right)  },\Phi^{\left(  1\right)  }\right\}
=O\left(  \Phi^{\left(  ...i\right)  }\right)
\end{array}
\right\}  ,\;i=1,...,\aleph_{\varphi}-1,\;s=i+1,...,\aleph_{\varphi
}\,,\nonumber\\
&  \left\{  \chi^{\left(  i|a\right)  },\bar{H}+\epsilon\right\}
=\chi^{\left(  i+1|a\right)  }+O\left(  \Phi^{\left(  ...i\right)  }\right)
\,,\;i=1,...,\aleph_{\chi}-1\,,\;a=i+1,...,\aleph_{\chi}\,,\nonumber\\
&  \left\{  \chi^{\left(  i|i\right)  },\bar{H}+\epsilon\right\}  =O\left(
\Phi^{\left(  ...i\right)  }\right)  \,,\;i=1,...,\aleph_{\chi}\,,\nonumber\\
&  \left\{  \chi^{\left(  i|a\right)  },\Phi^{\left(  1\right)  }\right\}
=O\left(  \Phi^{\left(  ...i\right)  }\right)  \,,\;i=1,...,\aleph_{\chi
}\,,\;a=1,...,\aleph_{\chi}\,,\nonumber\\
&  \left\{  \varphi^{\left(  i|s\right)  },\varphi^{\left(  j|v\right)
}\right\}  =\delta_{s,v}\delta_{i,s+1-j}\theta^{\left(  s\right)  }+O\left(
\Phi\right)  \,,\;\;i\leq s\,,\;j\leq v\,,\;\ s,v=1,...,\aleph_{\varphi}\,.
\label{2.6}%
\end{align}
Here, $\Phi^{\left(  ...i\right)  }=\left(  \Phi^{\left(  1\right)  }%
,...,\Phi^{\left(  i\right)  }\right)  ,$%
\begin{equation}
\bar{H}=H+\bar{\lambda}_{s}\varphi^{\left(  1\left|  s\right|  \right)  }\,,
\label{2.7}%
\end{equation}
where $\bar{\lambda}_{s}=\bar{\lambda}_{s}\left(  \eta\right)  $ are
expressions for the Lagrange multipliers to the primary SCC that are
determined by the Dirac procedure, and $\ \theta^{(s)}$ is a nonsingular
matrix. It should be noted that the Hamiltonian $\bar{H}$ differs from the
total Hamiltonian $H^{\left(  1\right)  }$ by terms quadratic in the extremals
and by FCC:%
\begin{equation}
H^{\left(  1\right)  }=\bar{H}+\Lambda_{s}\varphi^{\left(  1\left|  s\right|
\right)  }+\lambda^{a}\chi^{\left(  1\left|  a\right|  \right)  }%
\,,\;\Lambda_{s}=\lambda_{s}-\bar{\lambda}_{s}=O\left(  \frac{\delta
S_{\mathrm{H}}}{\delta\mathbf{\eta}}\right)  \,. \label{2.8}%
\end{equation}

From the properties of the orthogonal constraint basis, one can see that the
Poisson brackets of SCC from different chains vanish on the constraint
surface. Within the Dirac procedure, the group $\varphi^{\left(  1|s\right)
}$ of primary SCC produces SCC of the second stage, third stage, and so on,
which belong to the same chain,\ $\varphi^{\left(  1|s\right)  }%
\rightarrow\varphi^{\left(  2|s\right)  }\rightarrow\varphi^{\left(
3|s\right)  }\rightarrow\cdot\cdot\cdot\rightarrow\varphi^{\left(  s|s\right)
}.$\ The chain of SCC labeled by the number $s$\ ends with the group of the
$s$-th-stage constraints. The consistency conditions for the SCC
$\varphi^{\left(  i|i\right)  }$ of the $i$-th stage determine\ the Lagrange
multipliers $\lambda_{i}$\ to be $\bar{\lambda}_{i}\,.$ We stress that the
consistency conditions for the SCC $\varphi^{\left(  i|s\right)  },$\ $s>i$,
of the $i$-th stage produce SCC $\varphi^{\left(  i+1|s\right)  }$ of the
$i+1$-th stage.

At the same time, the group $\chi^{\left(  1|a\right)  }$\ of primary
FCC\ produces FCC of the second stage, third stage, and so on, which belong to
the same chain,$\;\chi^{\left(  1|a\right)  }\rightarrow\chi^{\left(
2|a\right)  }\rightarrow\chi^{\left(  3|a\right)  }\rightarrow\cdot\cdot
\cdot\rightarrow\chi^{\left(  a|a\right)  }$.\textbf{ }The consistency
conditions for the FCC $\chi^{\left(  i|s\right)  },$\ $s>i$, of the $i$-th
stage produce the FCC $\chi^{\left(  i+1|s\right)  }$ of the $i+1$-th stage.
The chain of FCC labeled by the number $a$\ ends with the group of the
$a$-th-stage constraints. The consistency conditions for the FCC
$\chi^{\left(  i|i\right)  }$ of the $i$-th stage do not produce any new
constraints and do not determine any Lagrange multipliers. Thus, the Lagrange
multipliers $\lambda_{\chi}$\ are not determined by the Dirac procedure (or by
the complete set of equations of motion).

The described hierarchy of constraints in the orthogonal basis (within the
Dirac procedure) looks schematically as follows:
\[%
\begin{array}
[c]{lllllllllll}%
\varphi^{\left(  1|1\right)  } & \rightarrow & \bar{\lambda}_{1} &  &  &  &  &
&  &  & \\
\varphi^{\left(  1|2\right)  } & \rightarrow & \varphi^{\left(  2|2\right)  }
& \rightarrow & \bar{\lambda}_{2} &  &  &  &  &  & \\
\vdots & \rightarrow & \vdots & \rightarrow & \vdots & \vdots & \vdots &  &  &
& \\
\varphi^{\left(  1|\aleph-1\right)  } & \rightarrow & \varphi^{\left(
2|\aleph-1\right)  } & \rightarrow & \varphi^{\left(  3|\aleph-1\right)  } &
\cdots & \varphi^{\left(  \aleph-1|\aleph-1\right)  } & \rightarrow &
\bar{\lambda}_{\aleph-1} &  & \\
\varphi^{\left(  1|\aleph\right)  } & \rightarrow & \varphi^{\left(
2|\aleph\right)  } & \rightarrow & \varphi^{\left(  3|\aleph\right)  } &
\cdots & \varphi^{\left(  \aleph-1|\aleph\right)  } & \rightarrow &
\varphi^{\left(  \aleph|\aleph\right)  } & \rightarrow & \bar{\lambda}%
_{\aleph}\\
\chi^{\left(  1|\aleph\right)  } & \rightarrow & \chi^{\left(  2|\aleph
\right)  } & \rightarrow & \chi^{\left(  3|\aleph\right)  } & \cdots &
\chi^{\left(  \aleph-1|\aleph\right)  } & \rightarrow & \chi^{\left(
\aleph|\aleph\right)  } & \rightarrow & O(\Phi)\\
\chi^{\left(  1|\aleph-1\right)  } & \rightarrow & \chi^{\left(
2|\aleph-1\right)  } & \rightarrow & \chi^{\left(  3|\aleph-1\right)  } &
\cdots & \chi^{\left(  \aleph-1|\aleph-1\right)  } & \rightarrow &
O(\Phi^{\left(  ...\aleph-1\right)  }) &  & \\
\vdots & \rightarrow & \vdots & \rightarrow & \vdots & \vdots & \vdots &  &  &
& \\
\chi^{\left(  1|2\right)  } & \rightarrow & \chi^{\left(  2|2\right)  } &
\rightarrow & O(\Phi^{\left(  ...2\right)  }) &  &  &  &  &  & \\
\chi^{\left(  1|1\right)  } & \rightarrow & O(\Phi^{\left(  1\right)  }) &  &
&  &  &  &  &  &
\end{array}
.
\]

We note that the properties of the orthogonal constraint basis are extremely
helpful in analyzing the symmetry equation; in particular, this equation in
the orthogonal basis can be solved algebraically (see below). The properties
of the basis allow one to conjecture (and then strictly prove) the form of the
conserved charges as decompositions in the orthogonal constraint basis. For
example, these properties imply that the SCC $\varphi^{\left(  i|i\right)  }$
cannot enter linearly into the conserved charges. At the same time, one can
see that only the FCC $\chi^{\left(  i|i\right)  }$ enter the gauge charges
multiplied by independent gauge parameters; other FCC $\chi^{\left(
i|a\right)  },\;a>i$, are multiplied by factors that must contain derivatives
of the same gauge parameters.

It is also important to recall that there exists a canonical transformation
from the initial phase-space variables $\eta$ to\ the special phase-space
variables $\vartheta=\left(  \omega,Q,\Omega\right)  $; see \cite{GitTy90}.
The variables $\omega$ are physical, and the variables $\Omega$ define the
constraint surface by the equations\ $\Omega=0$. The dynamics of the physical
variables $\omega$ is governed by the physical action $S_{\mathrm{Ph}}\left[
\omega\right]  ,$%
\begin{equation}
S_{\mathrm{Ph}}\left[  \omega\right]  =\left.  S_{\mathrm{H}}\right|
_{\Omega=0}\,. \label{2.9}%
\end{equation}
The variables $Q$ are nonphysical. In theories with FCC, the actions $S$ and
$S_{\mathrm{H}}$ are dynamically equivalent, while are both dynamically
nonequivalent to the physical action $S_{\mathrm{Ph}}$.

In what follows, we use a set of LF, $\Gamma$,%
\begin{equation}
\Gamma=(\Phi,J)=O\left(  \frac{\delta S_{\mathrm{H}}}{\delta\mathbf{\eta}%
}\right)  ,\;J=(\mathcal{I},\;\Lambda),\;\mathcal{I}=\dot{\eta}-\left\{
\eta,\bar{H}\right\}  -\left\{  \eta,\chi^{\left(  1\left|  a\right|  \right)
}\right\}  \lambda^{a}\,, \label{2.10}%
\end{equation}
for the complete set of extremals. Taking (\ref{2.1}) into account, one can
easily verify that the set $\Gamma$ is equivalent to the complete set of
extremals $\delta S_{\mathrm{H}}/\delta\mathbf{\eta}$. We also note that the
set of variables $\eta^{\lbrack]}$, $\lambda_{s}^{[]}$, $\lambda^{a[]}$ is
equivalent to the set $\eta$, $J^{[]}$, $\lambda^{a[]}$. Further, we often use
the latter variables to analyze the symmetry equation.

\section{Gauge symmetries}

Analyzing the symmetry equation, we prove below the following assertion.

In theories with FCC, there exist nontrivial symmetries $\delta_{\nu
}\mathbf{\eta}$, $G_{\nu}$ of the Hamiltonian action $S_{\mathrm{H}}$ that are
gauge transformations. These symmetries are parameterized by gauge parameters
$\nu$.\ These parameters are arbitrary functions of time\footnote{We have
included $t$ in the arguments of the functions $\nu$ explicitly, in contrast
to other cases, to emphasize this dependence.} and arbitrary LF of
$\mathbf{\eta}^{[]}$,\thinspace%
\[
\nu=\nu_{\left(  a\right)  \sigma_{a}}\left(  t,\mathbf{\eta}^{[]}\right)
\,,\;a=1,...,\aleph_{\chi}\,.
\]
The gauge parameters are labeled by an index $a$ (the number of the
corresponding FCC chain) and by a fine index\footnote{Sometimes, the fine
index $\sigma_{i}$ is omitted, but summation over the index $i$ always assumes
the summation over $\sigma_{i}$ as well.} $\sigma_{a}$ that labels FCC in the
chain $a.$ The complete number of all gauge parameters is equal to the number
of all primary FCC:%
\begin{equation}
\left[  \nu\right]  =\left[  \chi^{\left(  1\right)  }\right]  \,. \label{3.1}%
\end{equation}
The corresponding conserved charge (gauge charge) is an LF, $G_{\nu}=G_{\nu
}\left(  \eta,\lambda^{\lbrack]},\nu^{\lbrack]}\right)  $, which vanishes on
the extremals. The gauge charge has the following representation in terms of
the orthogonal constraint basis:%
\begin{equation}
G_{\nu}=\sum_{a=1}^{\aleph_{\chi}}\nu_{(a)}\chi^{\left(  a|a\right)  }%
+\sum_{i=1}^{\aleph_{\chi}-1}\sum_{a=i+1}^{\aleph_{\chi}}C_{i|a}^{\chi}%
\chi^{\left(  i|a\right)  }+\sum_{i=1}^{\aleph_{\chi}-1}\sum_{s=i+1}^{\aleph
}C_{i|s}^{\varphi}\varphi^{\left(  i|s\right)  }\,. \label{3.2}%
\end{equation}
Here, $C_{i|s}^{\varphi}=C_{i|s}^{\varphi}\left(  \eta,\lambda^{\lbrack]}%
,\nu^{\lbrack]}\right)  $\ and $C_{i|a}^{\chi}=C_{i|a}^{\chi}\left(
\eta,\lambda^{\lbrack]},\nu^{\lbrack]}\right)  $\ are some LF, which can be
determined from the symmetry equation in an algebraic way. The gauge charge
depends both on the gauge parameters and on their time derivatives up to the
order $\aleph_{\chi}-1$,%
\begin{equation}
G_{\nu}=\sum_{a=1}^{\aleph_{\chi}}\sum_{m=0}^{a-1}G_{m}^{a}(\eta
,\lambda^{\lbrack]})\nu_{(a)}^{[m]}\,. \label{3.3}%
\end{equation}
Here, $G_{m}^{a}(\eta,\lambda^{\lbrack]})$\ are some LF. The total number of
independent gauge parameters, together with their time derivatives that
essentially enter in the gauge charge, is equal to the number of all FCC:%
\begin{equation}
\sum_{m=0}[\nu^{\lbrack m]}]=\left[  \chi\right]  \,. \label{3.4}%
\end{equation}
The gauge charge is the generating function for the variations $\delta\eta
$\ of the phase-space variables:%
\begin{equation}
\delta_{\nu}\eta=\left\{  \eta,G_{\nu}\right\}  =\left\{  \eta,\eta
^{A}\right\}  \frac{\partial G_{\nu}}{\partial\eta^{A}}\,. \label{3.5}%
\end{equation}
(Note that the Poisson bracket in (\ref{3.5}) acts only on the explicit
dependence of the gauge charge of $\eta$ .) Therefore, the total number of
independent gauge parameters, together with their time derivatives that
essentially enter in the variations $\delta_{\nu}\eta$,\ is also equal to the
number of all FCC. The variations $\delta_{\nu}\lambda$\ contain an additional
time derivative of the gauge parameters, namely, they have the form%
\begin{equation}
\delta_{\nu}\lambda=\sum_{a=1}^{\aleph_{\chi}}\sum_{m=0}^{a}\Upsilon_{m}%
^{a}\,\nu_{(a)}^{[m]}\,, \label{3.6}%
\end{equation}
where $\Upsilon_{m}^{a}=\Upsilon_{m}^{a}(\eta,\lambda^{\lbrack]})$\ are some LF.

The above assertion is proved below. At the same time, we present a
constructive procedure of finding the gauge transformations, based on solving
the symmetry equation in terms of the orthogonal constraint basis.

Consider the symmetry equation (\ref{2.2}). Taking into account the action
structure (\ref{2.1}) and the anticipated form (\ref{3.5}) of the variations
$\delta\eta,$ we rewrite this equation as follows:%
\begin{equation}
\hat{H}G-\lambda^{a}\left\{  \chi^{(1\left|  a\right|  )},G\right\}
-\Lambda_{s}\{\varphi^{\left(  1\left|  s\right|  \right)  },G\}=\,\varphi
^{(1\left|  s\right|  )}\hat{\delta}\Lambda_{s}+\chi^{\left(  1\left|
a\right|  \right)  }\delta\lambda^{a}\,,\;\;\hat{\delta}\Lambda_{s}%
=\delta\lambda_{s}-\{\bar{\lambda}_{s},G\}\,, \label{3.7}%
\end{equation}
where the operator $\hat{H}$ is defined on any LF $F(\eta,\lambda^{\lbrack
]},\nu^{\lbrack]})$ as%
\begin{equation}
\hat{H}F=\left\{  F,\bar{H}+\epsilon\right\}  +\frac{\partial F}%
{\partial\lambda^{\left[  m\right]  }}\lambda^{\left[  m+1\right]
}+\frac{\partial F}{\partial\nu^{\left[  m\right]  }}\nu^{\left[  m+1\right]
}\,. \label{3.8}%
\end{equation}

Let us try to solve equation (\ref{3.7}) with the gauge charge $G_{\nu}$ of
the form (\ref{3.2}). Thus, we obtain the following equation for the LF
$\mathbf{C}_{i}=\left(  C_{i|s}^{\varphi}\,,\,C_{i|a}^{\chi}\right)  $:%
\begin{align}
&  \sum_{i=1}^{\aleph_{\chi}-1}\left(  \mathbf{C}_{i}\left[  \Phi^{\left(
i+1\right)  }+O\left(  \Phi^{\left(  ...i\right)  }\right)  \right]
+\Phi^{\left(  i\right)  }\left[  \hat{H}\mathbf{C}_{i}-\lambda^{a}\left\{
\chi^{(1\left|  a\right|  )},\mathbf{C}_{i}\right\}  -\Lambda_{s}%
\{\varphi^{\left(  1\left|  s\right|  \right)  },\mathbf{C}_{i}\}\right]
\right) \nonumber\\
&  +\Pi_{\aleph_{\chi}}=\varphi^{\left(  1\left|  s\right|  \right)  }%
\hat{\delta}\Lambda_{s}+\chi^{\left(  1\left|  a\right|  \right)  }%
\delta\lambda^{a}\,, \label{3.9}%
\end{align}
where%
\[
\Pi_{\aleph_{\chi}}=\sum_{a=1}^{\aleph_{\chi}}\left[  Z^{a}+\chi^{\left(
a|a\right)  }d_{t}\right]  \nu_{(a)}\,,\;Z^{a}=\hat{H}\chi^{\left(
a|a\right)  }-\lambda^{b}\left\{  \chi^{(1\left|  b\right|  )},\chi^{\left(
a|a\right)  }\right\}  -\Lambda_{s}\{\varphi^{\left(  1\left|  s\right|
\right)  },\chi^{\left(  a|a\right)  }\}=O\left(  \Phi^{\left(  ...a\right)
}\right)  \,.
\]
Considering equation (\ref{3.9}) on the constraint surface $\Phi^{\left(
...\aleph_{\chi}-1\right)  }=0,$ we obtain
\[
\mathbf{C}_{\aleph_{\chi}-1}\Phi^{\left(  \aleph_{\chi}\right)  }+\nu
_{(\aleph_{\chi})}\kappa\Phi^{\left(  \aleph_{\chi}\right)  }+\nu
_{(\aleph_{\chi})}^{\left[  1\right]  }\chi^{\left(  \aleph_{\chi}%
|\aleph_{\chi}\right)  }=0
\]
with allowance made for the relation%
\begin{equation}
\left.  Z^{\aleph_{\chi}}\right|  _{\Phi^{\left(  ...\aleph_{\chi}-1\right)
}=0}=O\left(  \Phi^{\left(  \aleph_{\chi}\right)  }\right)  =\kappa
\Phi^{\left(  \aleph_{\chi}\right)  }\,. \label{3.10}%
\end{equation}
where $\kappa$ is a matrix with elements that are LF. Therefore, we can
determine $\mathbf{C}_{\aleph_{\chi}-1}$ as linear combinations of
$\nu_{(\aleph_{\chi})}$ and $\nu_{(\aleph_{\chi})}^{\left[  1\right]  }$ to
obey the latter equation. Substituting thus determined $\mathbf{C}%
_{\aleph_{\chi}-1}$ into (\ref{3.9}), we arrive at the equation%
\begin{align}
&  \sum_{i=1}^{\aleph_{\chi}-2}\left[  \mathbf{C}_{i}[\Phi^{\left(
i+1\right)  }+O\left(  \Phi^{\left(  ...i\right)  }\right)  ]+\Phi^{\left(
i\right)  }[\hat{H}\mathbf{C}_{i}-\lambda^{a}\left\{  \chi^{(1\left|
a\right|  )},\mathbf{C}_{i}\right\}  -\Lambda_{s}\{\varphi^{\left(  1\left|
s\right|  \right)  },\mathbf{C}_{i}\}]\right] \nonumber\\
&  +\Pi_{\aleph_{\chi}-1}=\varphi^{\left(  1\left|  s\right|  \right)  }%
\hat{\delta}\Lambda_{s}+\chi^{\left(  1\left|  a\right|  \right)  }%
\delta\lambda^{a}\,, \label{3.11}%
\end{align}
where%
\[
\Pi_{\aleph_{\chi}-1}=\sum_{a=1}^{\aleph_{\chi}-1}\left[  Z^{a}+\chi^{\left(
a|a\right)  }d_{t}\right]  \nu_{(a)}+\sum_{k=0}^{2}\phi_{k}\nu_{(\aleph_{\chi
})}^{\left[  k\right]  }\,,\;\;\phi_{k}=O\left(  \Phi^{\left(  ...\aleph
_{\chi}-1\right)  }\right)  \,.
\]
Considering equation (\ref{3.11}) on the constraint surface $\Phi^{\left(
...\aleph_{\chi}-2\right)  }=0$, we obtain%
\begin{align}
&  \mathbf{C}_{\aleph_{\chi}-2}\Phi^{\left(  \aleph_{\chi}-1\right)  }%
+\nu_{(\aleph_{\chi}-1)}^{\left[  1\right]  }\chi^{\left(  \aleph_{\chi
}-1|\aleph_{\chi}-1\right)  }\nonumber\\
&  \,+\left.  \left(  \nu_{(\aleph_{\chi}-1)}Z^{\aleph_{\chi}-1}+\sum
_{k=0}^{2}\phi_{k}\nu_{(\aleph_{\chi})}^{\left[  k\right]  }\right)  \right|
_{\Phi^{\left(  ...\aleph_{\chi}-2\right)  }=0}=0\,. \label{3.12}%
\end{align}
By analogy with (\ref{3.10}), we find that all matrix LF $Z^{\aleph_{\chi}-1}$
and $\phi_{k}$ are proportional to the constraints $\Phi^{\left(  \aleph
_{\chi}-1\right)  }$. Therefore, we can determine $\mathbf{C}_{\aleph_{\chi
}-2}$ as linear combinations of $\nu_{(\aleph_{\chi}-1)}$, $\nu_{(\aleph
_{\chi}-1)}^{\left[  1\right]  }$, and $\nu_{(\aleph_{\chi})}$, $\nu
_{(\aleph_{\chi})}^{\left[  1\right]  }$, $\nu_{(\aleph_{\chi})}^{\left[
2\right]  }$ to obey Eq. (\ref{3.12}). Proceeding in the same manner, we
determine any $\mathbf{C}_{i}$ as the following linear combinations of the
gauge parameters and their time derivatives:%
\begin{equation}
\mathbf{C}_{i}=\sum_{a=i+1}^{\aleph_{\chi}}\sum_{m=0}^{a-i}S_{im}^{a}\nu
_{(a)}^{[m]}\,. \label{3.13}%
\end{equation}
Here, $S_{im}^{a}=S_{im}^{a}\left(  \eta,\lambda^{\lbrack]}\right)  $ are some LF.

In addition, we can see that%
\begin{equation}
C^{\varphi}=O(\Gamma)\,. \label{3.14}%
\end{equation}
Indeed, Eq. (\ref{3.14}) follows from the relations (we recall that symmetry
variations of extremals are proportional to extremals)%
\[
\hat{\delta}\varphi^{(s+1-i|s)}=O(\Gamma)=\left\{  \varphi^{(s+1-i|s)},G_{\nu
}\right\}  =C_{i|s}^{\varphi}\left\{  \varphi^{(s+1-i|s)},\varphi^{\left(
i|s\right)  }\right\}  +O(\Phi)\,,
\]
with allowance made for (\ref{2.6}).

Finally, we conclude that the gauge charge $G_{\nu}$ has the following
representation:%
\begin{equation}
G_{\nu}=\sum_{m=1}^{\aleph_{\chi}}\sum_{a=m}^{\aleph_{\chi}}G^{ma}\nu
_{(a)}^{[m-1]}\,. \label{3.15}%
\end{equation}
Here, $G^{ma}=G^{ma}(\eta,\lambda^{\lbrack]})$ are LF of the form%
\begin{equation}
G^{ma}=\sum_{k=1}^{\aleph_{\chi}}\sum_{b=k}^{\aleph_{\chi}}\chi^{\left(
k|b\right)  }\mathcal{C}_{kb}^{ma}+O\left(  \Gamma^{2}\right)  =O\left(
\chi\right)  +O\left(  \Gamma^{2}\right)  \,, \label{3.16}%
\end{equation}
where $\mathcal{C}_{kb}^{ma}=\mathcal{C}_{kb}^{ma}(\eta,\lambda^{\lbrack]})$
are some LF. Then the form of the variations $\delta_{\nu}\eta$ follows from
(\ref{3.5}),%
\begin{equation}
\delta_{\nu}\eta=\left(  \sum_{k=1}^{\aleph_{\chi}}\sum_{b=k}^{\aleph_{\chi}%
}\right)  \left(  \sum_{m=1}^{\aleph_{\chi}}\sum_{a=m}^{\aleph_{\chi}}\right)
\left\{  \eta,\chi^{\left(  k|b\right)  }\right\}  \mathcal{C}_{kb}^{ma}%
\nu_{(a)}^{[m-1]}+O\left(  \Gamma\right)  \,. \label{3.18}%
\end{equation}

After substituting the obtained functions $\mathbf{C}_{i}$ back into equation
(\ref{3.9}), its left-hand side turns out to be proportional to primary
constraints. Since these constraints are linearly independent by construction,
we can find all the variations $\delta_{\nu}\lambda$ from this equation. Their
general structure is given by Eq. (\ref{3.6}). In particular, one can see that%
\begin{equation}
\delta_{\nu}\lambda^{a}=\sum_{b=1}^{\aleph_{\chi}}\mathcal{D}^{ab}\nu
_{(b)}^{[b]}+O(\nu_{(j)}^{[l]}\,,\;l<j)\,, \label{3.19}%
\end{equation}
where $\mathcal{D}^{ab}=\mathcal{D}^{ab}(\eta,\lambda^{\lbrack]})$ are some
LF. Note that the LF $G^{ma}$ and $\Upsilon_{m}^{a}$ (as well as
$\mathcal{C}_{kb}^{ma}$ and $\mathcal{D}^{ab}$) do not depend on the gauge
parameters and are, in this sense, universal.

The matrices $\mathcal{C}$ and $\mathcal{D}$ are nonsingular. Indeed, as was
demonstrated in \cite{218}, they are constant matrices of the form%
\[
\mathcal{C}_{ka}^{mb}=\delta_{b,a}\delta_{m,b-k+1}\,,\;\mathcal{D}^{ab}%
=\delta_{a,b}\,
\]
in the quadratic approximation. Since these matrices are nonsingular in the
quadratic approximation, they are nonsingular in the exact perturbative theory
at least in a vicinity of the consideration point, which can be chosen as the
zero point of the jet space of $\eta^{\lbrack]}.$

The gauge charge and the variations $\delta_{\nu}\eta$ essentially depend on
all gauge parameters and their time derivatives $\nu_{(a)}^{\left[  m\right]
}$, $m=0,...,a-1$,$\;a=1,...,\aleph_{\chi}\,,$ while the variations
$\delta_{\nu}\lambda^{a}$ essentially depend on all derivatives $\nu
_{(a)}^{\left[  a\right]  }$. Indeed,%
\begin{align*}
&  \frac{\partial G_{\nu}}{\partial\nu_{(a)}^{[m]}}=\mathcal{P}^{(a-m|a)}%
+O\left(  \eta^{2}\right)  \,,\;\;m=0,...,a-1\,,\\
&  \frac{\partial\left(  \delta_{\nu}\lambda^{a}\right)  }{\partial\nu
_{(b)}^{[b]}}=\delta_{a,b}+O\left(  \eta\right)  \,,\;a,b=1,...,\aleph_{\chi
}\,.
\end{align*}

As one can see from the structure of the gauge charge (in the special phase
space variables $\vartheta=\left(  \omega,Q,\Omega\right)  $; see
\cite{GitTy90}), the gauge transformations, taken on the extremals, transform
only the nonphysical variables $Q$ and $\lambda^{a}\,.$

Below, we are going to study the structure of arbitrary symmetry of the
general Hamiltonian action $S_{\mathrm{H}}$. In particular, we prove that,
with accuracy up to a trivial transformation, any gauge transformation can be
represented in the form (\ref{3.5}), (\ref{3.6}), with the gauge charge
(\ref{3.2}).

\section{Structure of arbitrary symmetry}

We prove below the following assertion.

Any symmetry $\delta\mathbf{\eta}$, $G$\ of the Hamiltonian action
$S_{\mathrm{H}}$\ can be represented as the sum of three types of symmetries,%
\begin{equation}
\left(
\begin{array}
[c]{c}%
\delta\mathbf{\eta}\\
G
\end{array}
\right)  =\left(
\begin{array}
[c]{c}%
\delta_{c}\mathbf{\eta}\\
G_{c}%
\end{array}
\right)  +\left(
\begin{array}
[c]{c}%
\delta_{\bar{\nu}}\mathbf{\eta}\\
G_{\bar{\nu}}%
\end{array}
\right)  +\left(
\begin{array}
[c]{c}%
\delta_{\mathrm{tr}}\mathbf{\eta}\\
G_{\mathrm{tr}}%
\end{array}
\right)  \,, \label{sdc.24}%
\end{equation}
such that:

The set $\delta_{c}\mathbf{\eta}$, $G_{c}$ is a global symmetry, canonical for
the phase-space variables $\eta$. The corresponding conserved charge $G_{c}%
$\ does not vanish on the extremals.

The set $\delta_{\bar{\nu}}\mathbf{\eta}$, $G_{\bar{\nu}}$\ is a gauge
transformation presented in the previous section, with fixed gauge parameters
(i.e., with a specific form of the functions $\nu=\bar{\nu}\left(
t,\mathbf{\eta}^{[]}\right)  $) that do not vanish on the extremals (in what
follows, we call such a symmetry a particular gauge transformation). The
corresponding conserved charge $G_{\bar{\nu}}$ vanishes on the extremals,
whereas the variations $\delta_{\bar{\nu}}\mathbf{\eta}$ do not.

The set $\delta_{\mathrm{tr}}\mathbf{\eta}$, $G_{\mathrm{tr}}$\ is a trivial
symmetry. All the variations $\delta_{\mathrm{tr}}\mathbf{\eta}$ and the
corresponding conserved charge $G_{\mathrm{tr}}$ vanish on the extremals. The
gauge charge $G_{\mathrm{tr}}$\ depends on the extremals as $G_{\mathrm{tr}%
}=O\left(  \Gamma^{2}\right)  .$

Below, we prove the above assertions and present a constructive way of finding
the components of the decomposition (\ref{sdc.24}). The procedure can be
divided into the four steps:

\subsection{Constructing the function\ $G_{J}\left(  \eta\right)  $}

Supposing that $\delta\mathbf{\eta}$, $G$ is a symmetry, and taking into
account the structure of the total Hamiltonian (\ref{2.8}), we can write the
symmetry equation (\ref{2.2}) as%
\begin{equation}
\delta\eta E^{-1}\mathcal{I}-\varphi^{\left(  1\left|  s\right|  \right)
}\hat{\delta}\Lambda_{s}-\Lambda_{s}\hat{\delta}\varphi^{\left(  1\left|
s\right|  \right)  }-\chi^{(1\left|  a\right|  )}\delta\lambda^{a}%
+d_{t}G=0\,,\;\hat{\delta}\Lambda_{s}=\delta\lambda_{s}-\frac{\partial
\bar{\lambda}_{s}}{\partial\eta}\delta\eta\,. \label{sdc.26}%
\end{equation}
In our consideration, we use the set of variables $\eta$, $J^{[]}$,
$\lambda^{a[]},$ see sec. 3. We denote via $\delta_{J}\mathbf{\eta}$,
$G_{J}^{\prime}$ the corresponding zero-order terms in the decomposition of
the quantities $\delta\mathbf{\eta}$, $G$ with respect to the extremals
$J^{[]},$%
\begin{equation}
\left(
\begin{array}
[c]{c}%
\delta\mathbf{\eta}\\
G
\end{array}
\right)  =\left(
\begin{array}
[c]{l}%
\delta_{J}\mathbf{\eta}+O(J)\\
G_{J}^{\prime}+B_{m}J^{[m]}+O(J^{2})
\end{array}
\right)  , \label{sdc.25}%
\end{equation}
where $\delta_{J}\mathbf{\eta=}\delta_{J}\mathbf{\eta}(\eta,\lambda
^{a[]}),\;G_{J}^{\prime}=G_{J}^{\prime}(\eta,\lambda^{a[]})$, and $B_{m}%
=B_{m}(\eta,\lambda^{a[]}).$ We then rewrite equation (\ref{sdc.26}),
retaining only the terms of zero and first order with respect to the extremals
$J^{[]}$. Such an equation has the form%
\begin{align}
&  \delta_{J}\eta E^{-1}\mathcal{I}-\chi^{(1\left|  a\right|  )}\delta
_{J}\lambda^{a}=-\hat{H}G_{J}^{\prime}+\left\{  \chi^{(1\left|  a\right|
)},G_{J}^{\prime}\right\}  \lambda^{a}+\Lambda_{s}\{\varphi^{\left(  1\left|
s\right|  \right)  },G_{J}^{\prime}\}\nonumber\\
&  +\left\{  \eta,G_{J}^{\prime}\right\}  E^{-1}\mathcal{I}-J^{[m]}%
\hat{H}B_{m}+\lambda^{a}\left\{  \chi^{(1\left|  a\right|  )},B_{m}\right\}
J^{[m]}-B_{m}J^{[m+1]}+O(\Phi\Gamma)\,. \label{sdc.27}%
\end{align}
Contributions from the terms $\varphi^{\left(  1\left|  s\right|  \right)
}\hat{\delta}\Lambda_{s}$ and $\Lambda_{s}\hat{\delta}\varphi^{\left(
1\left|  s\right|  \right)  }$ are accumulated in $O(\Phi\Gamma)$, and the
operator $\hat{H}$ is defined by (\ref{3.8}).

Analyzing the terms with the extremals $J^{[]}$ (starting from the highest
derivative) in Eq. (\ref{sdc.27}), we can see that $B_{m}=O(\Phi)$ for every
$m$. Considering then the terms proportional to $\mathcal{I}$ in Eq.
(\ref{sdc.27}), we get the following expression:%
\[
\delta_{J}\eta=\left\{  \eta,G_{J}^{\prime}\right\}  +O(\Phi)
\]
for the variations $\delta_{J}\eta.$ We then see that%
\begin{equation}
\left\{  \Phi,G_{J}^{\prime}\right\}  =O(\Phi) \label{sdc.28}%
\end{equation}
with allowance made for the relations%
\[
\hat{\delta}\Phi=O(\Gamma)\Longrightarrow\hat{\delta}_{J}\Phi=O(\Phi)=\left\{
\Phi,G_{J}^{\prime}\right\}  +O(\Phi)\,.
\]
We can check that $\left\{  \chi,G_{J}^{\prime}\right\}  $ are first-class
functions, which implies that%
\begin{equation}
\left\{  \chi,G_{J}^{\prime}\right\}  =O(\chi)+O(\Phi^{2})\,. \label{sdc.29}%
\end{equation}

Considering the remaining terms in Eq. (\ref{sdc.27}), we get the equation%
\begin{equation}
\chi^{(1\left|  a\right|  )}\delta_{J}\lambda^{a}=\hat{H}G_{J}^{\prime
}+\lambda^{a}\left\{  \chi^{(1\left|  a\right|  )},G_{J}^{\prime}\right\}
+O(\Phi^{2})\,, \label{sdc.30}%
\end{equation}
which relates $\delta_{J}\lambda^{a}$ and $G_{J}^{\prime}$. This equation
allows us to study the function $G_{J}^{\prime}$ in more detail. To this and,
we rewrite the equation\ as%
\[
\left\{  G_{J}^{\prime},\bar{H}+\epsilon\right\}  +\frac{\partial
G_{J}^{\prime}}{\partial\lambda^{a\left[  m\right]  }}\lambda^{a\left[
m+1\right]  }=O(\chi)+O(\Phi^{2})\,,
\]
taking into account (\ref{3.8}) and (\ref{sdc.29}). Analyzing terms with the
Lagrange multipliers $\lambda_{\chi}^{[]}$ (starting from the highest
derivatives) in this equation, we can see that these multipliers can enter
only the terms that vanish on the constraint surface. For example, considering
the terms with the highest derivative $\lambda_{\chi}^{\left[  M+1\right]  }$
in the latter equation, we obtain%
\[
\frac{\partial G_{J}^{\prime}}{\partial\lambda^{a\left[  M\right]  }}%
=O(\chi)+O(\Phi^{2})\Longrightarrow G_{J}^{\prime}=G_{J}^{\prime}%
(\cdots\lambda^{a\left[  M-1\right]  })+O(\chi)+O(\Phi^{2})\,.
\]
Similarly, we finally conclude that $G_{J}^{\prime}$ has the structure%
\begin{equation}
G_{J}^{\prime}=G_{J}(\eta)+O(\chi)+O(\Phi^{2})=G_{J}(\eta)+B_{\chi}\chi
+O(\Phi^{2})\,, \label{sdc.30a}%
\end{equation}
where $B_{\chi}=B_{\chi}(\eta,\lambda^{a[]}).$ For the function $G_{J}\,,$ we
obtain%
\begin{equation}
\left\{  \varphi,G_{J}\right\}  =O(\Phi)\,,\;\left\{  \chi,G_{J}\right\}
=O(\chi)+O(\Phi^{2}) \label{sdc.31}%
\end{equation}
with allowance made for (\ref{sdc.28}) and (\ref{sdc.29}).

The relation
\[
\delta\Lambda=O(\Gamma)\Longrightarrow\delta_{J}\lambda_{s}-\left\{
\bar{\lambda}_{s},G_{J}+B_{\chi}\chi\right\}  =O(\Gamma)=O(\Phi)\,,
\]
defines the variations $\delta_{J}\lambda_{s}$, taken in the lowest order with
respect to the extremals.

In addition to relations (\ref{sdc.31}), the function $G_{J}(\eta)$ also
obeys
\begin{equation}
\left\{  G_{J},\bar{H}+\epsilon\right\}  =O(\chi)+O(\Phi^{2}) \label{sdc.31a}%
\end{equation}

Thus, the above consideration allows one to represent a refined version of the
representation (\ref{sdc.25})%
\begin{equation}
\left(
\begin{array}
[c]{l}%
\delta\eta\\
\delta\lambda_{s}\\
\delta\lambda^{a}\\
G
\end{array}
\right)  =\left(
\begin{array}
[c]{l}%
\left\{  \eta,G_{J}+B_{\chi}\chi\right\}  +O(\Gamma)\\
\left\{  \bar{\lambda}_{s},G_{J}+B_{\chi}\chi\right\}  +O(\Gamma)\\
\delta_{J}\lambda^{a}+O(J)\\
G_{J}+B_{\chi}\chi+O(\Gamma^{2})
\end{array}
\right)  , \label{sdc.37}%
\end{equation}
where $\delta_{J}\lambda^{a}=\delta_{J}\lambda^{a}(\eta,\lambda^{b[]}).$

\subsection{Constructing the function $G_{\Gamma}\left(  \eta\right)  $}

We now select from the function $G_{J}$ a part $G_{\Gamma}$ that does not
vanish on the constraint surface:%
\begin{equation}
G_{J}\left(  \eta\right)  =G_{\Gamma}\left(  \eta\right)  +G_{1}\left(
\eta\right)  \,,\;\;G_{\Gamma}=\left.  G_{J}\right|  _{\Phi=0}\,\,.
\label{sdc.32}%
\end{equation}
The function $G_{1}$ vanishes on this surface. Certainly, this decomposition
is not unique. We fix the procedure of such a decomposition in the special
variables $\vartheta=(\omega,Q,\Omega).$ To this end, we write%
\begin{equation}
G_{J}(\eta)=\tilde{G}_{J}\left(  \vartheta\right)  =g\left(  \omega\right)
+g_{1}\left(  \omega,Q\right)  Q+O\left(  \Omega\right)  \,. \label{sdc.33}%
\end{equation}
Considering the second relation of (\ref{sdc.31}) in the special variables, we
can see that the function $g_{1}\left(  \omega,Q\right)  $ must be zero.
Moreover, the last term $O\left(  \Omega\right)  $ in (\ref{sdc.33}) can be
specified to have the form $O\left(  \Omega\right)  =O\left(  \mathcal{P}%
\right)  +O\left(  \Omega^{2}\right)  $. We define $G_{\Gamma}\left(
\eta\right)  $ as%
\begin{equation}
G_{\Gamma}\left(  \eta\right)  =g\left(  \omega\right)  \,. \label{sdc.34}%
\end{equation}
Then,%
\[
G_{J}(\eta)=G_{\Gamma}(\eta)+G_{1}(\eta),\;G_{1}(\eta)=O(\chi)+O(\Phi^{2})\,,
\]
and, therefore,%
\begin{align}
&  G=G_{\Gamma}(\eta)+O(\chi)+O(\Gamma^{2})\,,\nonumber\\
&  \hat{H}G_{\Gamma}=O(\Phi),\;\left\{  \varphi,G_{\Gamma}\right\}
=O(\Phi),\;\left\{  \chi,G_{\Gamma}\right\}  =O(\chi)+O(\Phi^{2})\,,
\label{sdc.35}%
\end{align}
in virtue of (\ref{sdc.31}).

We now define the variations $\delta_{\Gamma}\mathbf{\eta}$ as%
\begin{equation}
\delta_{\Gamma}\eta=\left\{  \eta,G_{\Gamma}\right\}  ,\;\delta_{\Gamma
}\lambda_{s}=\left\{  \bar{\lambda}_{s},G_{\Gamma}\right\}  \,,\;\delta
_{\Gamma}\lambda^{a}=0\,. \label{sdc.36}%
\end{equation}

\subsection{Constructing the global canonical part of a symmetry}

The set $\delta_{\Gamma}\mathbf{\eta},$ $G_{\Gamma}$ is an approximate
symmetry of the action $S_{\mathrm{H}}$. Indeed, it obeys the equation%
\[
\delta_{\Gamma}\mathbf{\eta}\frac{\delta S_{\mathrm{H}}}{\delta\mathbf{\eta}%
}+d_{t}G_{\Gamma}=O(\chi)+O(\Gamma^{2})\,.
\]
However, this approximate symmetry can be modified to become an exact symmetry
$\delta_{c}\mathbf{\eta}$, $G_{c},$ such that%
\begin{equation}
\delta_{c}\eta=\left\{  \eta,G_{c}\right\}  \,,\;G_{c}=G_{\Gamma}+O\left(
\chi\right)  +O\left(  \Phi\Gamma\right)  \,. \label{sdc.38}%
\end{equation}
To justify this assertion, we are going, first of all, to demonstrate that the
symmetry equation has a solution with $G_{c}$ of the form%
\begin{align*}
&  G_{c}=G_{\Gamma}+\sum_{i=1}^{\aleph-1}\sum_{s=i+1}^{\aleph}C_{i|s}%
^{\varphi}\varphi^{\left(  i|s\right)  }+\sum_{i=1}^{\aleph-1}\sum
_{a=i+1}^{\aleph}C_{i|a}^{\chi}\chi^{\left(  i|a\right)  }\equiv G_{\Gamma
}+\sum_{i=1}^{\aleph-1}\mathbf{C}_{i}\Phi^{\left(  i\right)  }\,,\\
&  \mathbf{C}_{i}=\left(  C_{i|s}^{\varphi}\,,C_{i|a}^{\chi}\right)
=\mathbf{C}_{i}\left(  \eta,\lambda^{\lbrack]}\right)  \,.
\end{align*}

In the case under consideration, the symmetry equation (\ref{sdc.26}) can be
rewritten as%

\[
\hat{H}G_{c}-\lambda^{a}\left\{  \chi^{(1\left|  a\right|  )},G_{c}\right\}
-\Lambda_{s}\{\varphi^{\left(  1\left|  s\right|  \right)  },G_{c}%
\}=\varphi^{\left(  1\left|  s\right|  \right)  }\hat{\delta}_{c}\Lambda
_{s}+\chi^{\left(  1\left|  a\right|  \right)  }\hat{\delta}_{c}\lambda
^{a}\,,\;\;\hat{\delta}_{c}\Lambda_{s}=\delta_{c}\lambda_{s}-\{\bar{\lambda
}_{s},G_{c}\}\,,
\]
and then transformed to the form
\begin{align}
&  \sum_{i=1}^{\aleph-1}\left(  \mathbf{C}_{i}\left[  \Phi^{\left(
i+1\right)  }+O\left(  \Phi^{\left(  ...i\right)  }\right)  \right]  \right)
+\Phi^{\left(  i\right)  }\hat{H}\mathbf{C}_{i}+A=\varphi^{\left(  1\left|
s\right|  \right)  }\hat{\delta}_{c}\Lambda_{s}+\chi^{\left(  1\left|
a\right|  \right)  }\delta_{c}\lambda^{a}\,,\nonumber\\
&  A=\hat{H}G_{\Gamma}-\lambda^{a}\left\{  \chi^{(1\left|  a\right|
)},G_{\Gamma}\right\}  -\Lambda_{s}\{\varphi^{\left(  1\left|  s\right|
\right)  },G_{\Gamma}\}=O\left(  \chi\right)  +O\left(  \Phi^{2}\right)  \,,
\label{sdc.39}%
\end{align}
by using (\ref{sdc.35}). Equation (\ref{sdc.39}) can be analyzed by analogy
with equations (\ref{3.9}). For example, on the constraint surface
$\Phi^{\left(  ...\aleph-1\right)  }=0,$ we obtain
\[
C_{\aleph-1}\Phi^{\left(  \aleph\right)  }+a^{\left(  \aleph\right)  }%
\Phi^{\left(  \aleph\right)  }=0,\;\;a^{\left(  \aleph\right)  }\Phi^{\left(
\aleph\right)  }=\left.  A\right|  _{\Phi^{\left(  ...\aleph-1\right)  }}\,.
\]
We can choose $\mathbf{C}_{\aleph-1}=-a^{\left(  \aleph\right)  }.$ In the
same manner, we can find all the coefficients $\mathbf{C}$. After substituting
these coefficients back into equation (\ref{sdc.39}), we can see that its LHS
is proportional to $\Phi^{\left(  1\right)  }$, and, thus, the variations
$\delta_{c}\lambda$ can be found. The coefficients $\mathbf{C}$ and the
variations $\delta_{c}\mathbf{\eta}$ have the properties%
\[
\left.  \mathbf{C}_{i}\right|  _{G_{\Gamma}=0}=\left.  \delta_{c}\mathbf{\eta
}\right|  _{G_{\Gamma}=0}=0\,.
\]

In addition, the relation $\hat{\delta}_{c}\varphi=O(\Gamma)=\left\{
\varphi,G_{c}\right\}  $ implies $C^{\varphi}=O(\Gamma).$ Therefore, the
charge $G_{c}$\ has the structure%

\begin{equation}
G_{c}=G_{\Gamma}+O\left(  \chi\right)  +O\left(  \Phi\Gamma\right)  \,.
\label{sdc.40}%
\end{equation}
\emph{\ }

Note that, in general, the charge $G_{c}\ $depends on $\lambda^{\lbrack]}$ if
the Dirac procedure has more than two stages, i.e., $\aleph>2$; see examples
in the Discussion.

\subsection{Constructing the gauge and trivial parts of a symmetry}

At this stage, we represent the symmetry $\delta\mathbf{\eta}$, $G$ as%
\begin{equation}
\delta\mathbf{\eta}=\delta_{c}\mathbf{\eta}+\delta_{r}\mathbf{\eta
\,},\;G=G_{c}+G_{r}\,. \label{sdc.41}%
\end{equation}
Since $\delta_{c}\mathbf{\eta}$, $G_{c}$ is a symmetry, it is obvious that
$\delta_{r}\mathbf{\eta}$, $G_{r}$ is a symmetry as well. With the help of
Eqs. (\ref{sdc.31}) and (\ref{sdc.40}), we can prove the following relations:%
\begin{align}
&  G_{r}=\sum_{i=1}^{\aleph_{\chi}}\sum_{a=i}^{\aleph_{\chi}}K_{i|a}%
\chi^{\left(  i|a\right)  }\,+O\left(  \Gamma^{2}\right)  \,,\nonumber\\
&  \delta_{r}\eta=\sum_{i=1}^{\aleph_{\chi}}\sum_{a=i}^{\aleph_{\chi}}\left\{
\eta,\chi^{\left(  i|a\right)  }\right\}  K_{i|a}+O\left(  \Gamma\right)  \,,
\label{sdc.42}%
\end{align}
where $K_{i|a}=K_{i|a}\left(  \eta,\lambda_{\chi}^{[]}\right)  $ are some LF
that do not vanish on the constraint surface.

In turn, we represent the symmetry $\delta_{r}\mathbf{\eta}$, $G_{r}$ in the
form%
\begin{equation}
\delta_{r}\mathbf{\eta}=\delta_{\bar{\nu}}\mathbf{\eta}+\delta_{\mathrm{tr}%
}\mathbf{\eta}\,,\;G_{r}=G_{\bar{\nu}}+G_{\mathrm{tr}}\,, \label{sdc.43}%
\end{equation}
where the set $\delta_{\bar{\nu}}\mathbf{\eta}$, $G_{\bar{\nu}}$\ is a
particular gauge transformation with specific gauge parameters,
\begin{equation}
\nu_{(a)}=\bar{\nu}_{(a)}=K_{a|a}\,, \label{sdc.44}%
\end{equation}
which do not vanish on the constraint surface. This implies%
\begin{equation}
G_{\bar{\nu}}=O\left(  \chi\right)  +O\left(  \Gamma^{2}\right)
\,,\;\delta_{\bar{\nu}}\eta=\left\{  \eta,G_{\bar{\nu}}\right\}  \,.
\label{sdc.45}%
\end{equation}

From Eqs. (\ref{sdc.42}) it follows that $\delta_{\mathrm{tr}}\mathbf{\eta}$,
$G_{\mathrm{tr}}$ is a symmetry with the conserved charge of the form%
\[
G_{\mathrm{tr}}=G_{\mathrm{tr}}^{\prime}+O(\Gamma^{2})\,,\;G_{\mathrm{tr}%
}^{\prime}=\sum_{i=1}^{\aleph_{\chi}-1}\sum_{a=i+1}^{\aleph_{\chi}}K_{i|a}%
\chi^{\left(  i|a\right)  }\,.
\]
We can show that $\delta_{\mathrm{tr}}\mathbf{\eta}$, $G_{\mathrm{tr}}$ is a
trivial symmetry. To this end, we can write for the symmetry $\delta
_{\mathrm{tr}}\mathbf{\eta}$, $G_{\mathrm{tr}}$ the decomposition of the form
(\ref{sdc.25}), taking into account that $B_{m}=O(\Phi),$%
\begin{equation}
\left(
\begin{array}
[c]{c}%
\delta\mathbf{\eta}_{\mathrm{tr}}\\
G_{\mathrm{tr}}%
\end{array}
\right)  =\left(
\begin{array}
[c]{l}%
\delta_{\mathrm{tr}J}\mathbf{\eta}+O(J)\\
G_{\mathrm{tr}J}^{\prime}+O(J^{2})\,,\;\;G_{\mathrm{tr}J}^{\prime
}=G_{\mathrm{tr}}^{\prime}+O(\Phi^{2})
\end{array}
\right)  . \label{sdc.46}%
\end{equation}
Here, $\delta_{\mathrm{tr}J}\mathbf{\eta=}\delta_{\mathrm{tr}J}\mathbf{\eta
}(\eta,\lambda_{\chi}^{[]})$ and $G_{\mathrm{tr}J}^{\prime}=G_{\mathrm{tr}%
J}^{\prime}(\eta,\lambda_{\chi}^{[]}).\;$All the relations that take place for
the quantities $\delta_{J}\mathbf{\eta}$, $G_{J}$ hold for the quantities
$\delta_{\mathrm{tr}J}\mathbf{\eta}$, $G_{\mathrm{tr}J}^{\prime}$ as well. In
particular, the charge $G_{\mathrm{tr}}^{\prime}$ obeys the equation%
\begin{align}
&  \chi^{(1\left|  a\right|  )}\delta_{\mathrm{tr}}^{\prime}\lambda
^{a}=\hat{H}G_{\mathrm{tr}}^{\prime}+\lambda^{a}\left\{  \chi^{(1\left|
a\right|  )},G_{\mathrm{tr}}^{\prime}\right\}  +O(\Phi^{2})\,,\nonumber\\
&  \delta_{\mathrm{tr}}^{\prime}\lambda^{a}=\left.  \delta_{\mathrm{tr}%
}\lambda^{a}\right|  _{\Gamma=0}=\delta_{\mathrm{tr}J}\lambda^{a}+O(\Phi)\,,
\label{sdc.47}%
\end{align}
which is similar to (\ref{sdc.30}). Equation (\ref{sdc.47}) implies%
\[
\sum_{i=1}^{\aleph_{\chi}-1}\sum_{a=i+1}^{\aleph_{\chi}}\left(  \chi^{\left(
i|a\right)  }\hat{H}K_{i|a}+K_{i|a}[\chi^{\left(  i+1|a\right)  }%
+O(\Phi^{(...i)})]\right)  =\chi^{(1\left|  s\right|  )}\delta_{\mathrm{tr}%
}^{\prime}\lambda_{s}+O(\Phi^{2})
\]
for the LF $K_{i|a}\,,\;a=i+1,...,\aleph_{\chi}$. Considering the latter
equation on the constraint surface $\Phi^{\left(  ...\aleph_{\chi}-1\right)
}=0$, we obtain%
\[
K_{\aleph_{\chi}-1|\aleph_{\chi}}\chi^{\left(  \aleph_{\chi}|\aleph_{\chi
}\right)  }=O(\Phi^{2})\Longrightarrow K_{\aleph_{\chi}-1|\aleph_{\chi}%
}=O(\Phi)\,.
\]
Substituting the above expression for $K_{\aleph_{\chi}-1|\aleph_{\chi}}$ into
(\ref{sdc.47}) and considering the resulting equation on the constraint
surface $\Phi^{\left(  ...\aleph_{\chi}-2\right)  }=0,$ we obtain
$K_{\aleph_{\chi}-2|\aleph_{\chi}}=O(\Phi)$, and so on. Thus, we can see that
$K_{i|a}=O(\Phi)$, $a=i+1,...,\aleph_{\chi},$ and therefore%
\begin{equation}
G_{\mathrm{tr}}=O(\Gamma^{2})\,. \label{sdc.48}%
\end{equation}
It follows from (\ref{sdc.47}) that%
\begin{equation}
\chi^{(1\left|  a\right|  )}\delta_{\mathrm{tr}}^{\prime}\lambda^{a}%
=O(\Phi^{2})\Longrightarrow\delta_{\mathrm{tr}}^{\prime}\lambda^{a}%
=O(\Phi)\Longrightarrow\delta_{\mathrm{tr}}\lambda^{a}=O(\Gamma)\,.
\label{sdc.48a}%
\end{equation}

By construction, the transformation $\delta_{\mathrm{tr}J}$ is completely
similar to $\delta_{J}$. Therefore, relation (\ref{sdc.31}) holds for this
transformation and implies%
\[
\delta_{\mathrm{tr}J}\eta=\{\eta,G_{\mathrm{tr}}^{\prime}\}+O\left(
\Phi\right)  =O(\Phi)\,,\;\;\delta_{\mathrm{tr}J}\lambda_{s}=\{\bar{\lambda
}_{s},G_{\mathrm{tr}}^{\prime}\}+O\left(  \Phi\right)  =O(\Phi)\,.
\]
Therefore,%
\begin{equation}
\delta_{\mathrm{tr}}\eta=O(\Gamma)\,,\;\;\delta_{\mathrm{tr}}\lambda
_{s}=O(\Gamma)\,. \label{sdc.49}%
\end{equation}
In the following section, we show that any symmetry of a Hamiltonian action
that vanishes on extremals is a trivial symmetry. Therefore, relations
(\ref{sdc.48})--(\ref{sdc.49}) prove that the symmetry $\delta_{\mathrm{tr}%
}\mathbf{\eta}$, $G_{\mathrm{tr}}$ is trivial.

We can also prove that the reduction of symmetry variations $\delta\omega$\ to
the extremals yields global canonical symmetries $\delta_{\mathrm{ph}}\omega$
of the physical action with the conserved charge $g(\omega)$ that is a
reduction of the complete conserved charge to the extremals. In addition, any
global canonical symmetry $\delta_{\mathrm{ph}}\omega$, $g(\omega)$ of the
physical action$\,$can be extended to a nontrivial global symmetry $\delta
_{c}\mathbf{\eta}$, $G_{c}$\ of the Hamiltonian action\textbf{ }%
$S_{\mathrm{H}}$. Therefore, classes of nontrivial global symmetries of a
singular action are isomorphic to classes of nontrivial global canonical
symmetries of the corresponding Hamiltonian action. At the same time, these
classes are isomorphic to classes of nontrivial global canonical symmetries of
the corresponding physical action.

\section{Trivial symmetries}

In this section, we are going to prove that any symmetry of the Hamiltonian
action that vanishes on the extremals is a trivial symmetry. First, we prove
this assertion for a nonsingular Hamiltonian action, and then for the general
singular case.

\subsection{Nonsingular case}

We recall that the Hamiltonian action $S_{\mathrm{H}}$ of any nonsingular
theory has the canonical Hamiltonian form
\begin{equation}
S_{\mathrm{H}}\left[  \eta\right]  =\int\left[  p\dot{x}-H\left(  \eta\right)
\right]  dt\,,\;\frac{\delta S_{\mathrm{H}}}{\delta\eta}=0\Longrightarrow
\dot{\eta}=\{\eta,H\}\,. \label{sda.1}%
\end{equation}
Eqs. (\ref{sda.1}) follow from (\ref{2.1}) in the absence of constraints.
Below, we are going to prove that:

Symmetries of the canonical Hamiltonian action that vanish on the extremals
are trivial symmetries.

To prove this statement, we note that any symmetry $\delta\eta,\,G$ of the
Hamiltonian action $S_{\mathrm{H}}$\thinspace has to obey the symmetry
equation
\begin{equation}
\frac{\delta S_{\mathrm{H}}}{\delta\eta}\delta\eta+d_{t}G=0\,. \label{sda.3}%
\end{equation}

To analyze the symmetry equation (\ref{sda.3}), we are going to use, instead
of the variables $\eta^{\lbrack]}$, the equivalent set of variables
$\eta,\Gamma^{\lbrack]}$; see the comments in the end of sec. 3. Here,%
\begin{equation}
\Gamma_{A}=\frac{\delta S_{\mathrm{H}}}{\delta\eta^{A}}=E_{AB}^{-1}\dot{\eta
}^{B}-\frac{\partial H}{\partial\eta^{A}}\,,\;E^{AB}=\left\{  \eta^{A}%
,\eta^{B}\right\}  \,,\;E_{AB}^{-1}E^{BC}=\delta_{A}^{C}\,. \label{sda.7}%
\end{equation}
In terms of the new variables, the total time derivative of any LF $F\left(
\eta,\Gamma^{\lbrack]}\right)  $\ reads
\begin{equation}
\frac{dF}{dt}=\partial_{t}F+E^{AB}\left(  \Gamma_{B}+\frac{\partial
H}{\partial\eta^{B}}\right)  \frac{\partial F}{\partial\eta^{A}}+\sum
_{k=0}\Gamma_{A}^{\left[  k+1\right]  }\frac{\partial F}{\partial\Gamma
_{A}^{\left[  k\right]  }}\,, \label{sda.8}%
\end{equation}
and the symmetry equation takes the form%
\begin{equation}
\{G,H+\epsilon\}+\left(  \delta\eta^{A}+\frac{\partial G}{\partial\eta^{B}%
}E^{BA}\right)  \Gamma_{A}+\sum_{k=0}\Gamma_{A}^{\left[  k+1\right]
}\frac{\partial G}{\partial\Gamma_{A}^{\left[  k\right]  }}=0\,. \label{sda.5}%
\end{equation}
\ \ 

Let us now suppose that a symmetry $\delta\eta$ vanishes on the extremals,
i.e., $\delta\eta=O\left(  \Gamma\right)  .$ Representing the charge $G$ as
\[
G=G_{0}+G_{1}\,,\;G_{0}=\left.  G\right|  _{\Gamma=0}=G_{0}\left(
\eta\right)  \,,\;G_{1}=\sum_{m=0}B_{m}^{A}\left(  \eta\right)  \Gamma
_{A}^{\left[  m\right]  }+O\left(  \Gamma^{2}\right)  \,,
\]
and considering the terms of zero and first order with respect to the
extremals only, we obtain from (\ref{sda.5})
\begin{equation}
\{G_{0},H+\epsilon\}=0\,,\;\;\frac{\partial G_{0}}{\partial\eta^{B}}%
E^{BA}\Gamma_{A}+\sum_{m=0}\left[  \left\{  B_{m}^{A}\,,H+\epsilon\right\}
-B_{m}^{A}d_{t}\right]  \Gamma_{A}^{\left[  m\right]  }=0\,. \label{sda.6}%
\end{equation}
Analyzing the terms with the extremals $\Gamma^{\lbrack]}$ (starting from the
highest time derivative) in the second equation (\ref{sda.6}), we can verify
that all functions $B_{m}^{A}$ are zero. This fact implies that $\partial
_{t}G_{0}=0.$ Indeed,%
\[
B_{m}^{A}=0\Longrightarrow\frac{\partial G_{0}}{\partial\eta^{B}%
}=0\Longrightarrow\{G_{0},H+\epsilon\}=\{G_{0},\epsilon\}=\partial_{t}%
G_{0}=0\,.
\]
Taking into account that, in fact, the charge $G$ is defined up to a constant,
we can assume $G_{0}\equiv0.$ Thus, for any symmetry $\delta\eta$ that
vanishes on the extremals, the corresponding conserved charge also vanishes on
the extremals and has the form $G=G_{1}=O\left(  \Gamma^{2}\right)  .$

Let us represent this charge as follows:%
\begin{equation}
G=\sum_{m=0}^{N<\infty}g_{m}^{A}\Gamma_{A}^{[m]}\,, \label{sda.29}%
\end{equation}
where $g_{m}^{A}=g_{m}^{A}\left(  \eta,\Gamma^{\lbrack]}\right)  $ are some
LF. Substituting the representation (\ref{sda.29}) into (\ref{sda.5}), we
obtain the equation%
\begin{equation}
\sum_{m=0}^{N+1}f_{m}^{A}\Gamma_{A}^{\left[  m\right]  }=0\,, \label{sda.30}%
\end{equation}
where%
\begin{equation}
f_{0}^{A}=\delta\eta^{A}+d_{t}g_{0}^{A}\,;\;\;f_{m}^{A}=g_{m-1}^{A}+d_{t}%
g_{m}^{A}\,,\;\;m=1,...,N\,;\;\;f_{N+1}^{A}=g_{N}^{A}\,. \label{sda.31}%
\end{equation}
The general solution of equation (\ref{sda.30}) reads
\begin{equation}
f_{k}^{A}=\sum_{m=0}^{N+1}V_{k,m}^{AB}\Gamma_{B}^{\left[  m\right]  }\,,
\label{sda.32}%
\end{equation}
where LF $V_{k,m}^{AB}=V_{k,m}^{AB}\left(  \eta,\Gamma^{\lbrack]}\right)  $
are antisymmetric, $V_{m,k}^{BA}=-V_{k,m}^{AB}\,$. Relation (\ref{sda.32})
implies for the functions $g_{k}^{A}$ the following expressions:
\begin{equation}
g_{k}^{A}=-\sum_{s=0}^{N-k}\left(  -d_{t}\right)  ^{s}\sum_{m=0}%
^{N+1}V_{k+s+1,m}^{AB}\Gamma_{B}^{\left[  m\right]  }\,. \label{sda.33}%
\end{equation}
Then Eqs. (\ref{sda.31})-(\ref{sda.33}) determine $\delta\eta$ to be%
\begin{align}
&  \delta\eta^{A}=\hat{U}^{AB}\Gamma_{B}=\hat{U}^{AB}\frac{\delta
S_{\mathrm{H}}}{\delta\eta^{B}}\,,\nonumber\\
&  \hat{U}^{AB}=\sum_{m,k=0}^{N+1}\left(  -d_{t}\right)  ^{m}V_{m,k}%
^{AB}\left(  d_{t}\right)  ^{k}\,,\;\left(  \hat{U}^{T}\right)  ^{AB}%
=-\hat{U}^{AB}\,, \label{sda.34}%
\end{align}
where $\hat{U}^{AB}$ is an antisymmetric LO. Expression (\ref{sda.34}) implies
that $\delta\eta$ is a trivial symmetry of the action $S_{\mathrm{H}}$ .

\subsection{Singular case}

We consider here the general case of a singular theory. We are going to prove
that the symmetries of the corresponding Hamiltonian action that vanish on the
extremals are trivial symmetries.

First of all, we recall that:

a) The Hamiltonian action $S_{\mathrm{H}}\left[  \mathbf{\vartheta}\right]  $
of a singular theory (in the special phase-space variables $\vartheta=\left(
\omega,Q,\Omega\right)  )$ has the following structure:%
\begin{align}
&  S_{\mathrm{H}}\left[  \mathbf{\vartheta}\right]  =S_{\mathrm{ph}}\left[
\omega\right]  +S_{\mathrm{non-ph}}\left[  \mathbf{\vartheta}\right]
\,,\;\mathbf{\vartheta}=\left(  \vartheta,\lambda\right)  \,,\nonumber\\
&  S_{\mathrm{ph}}\left[  \omega\right]  =\int\left[  \omega_{p}\dot{\omega
}_{x}-H_{\mathrm{ph}}\left(  \omega\right)  \right]  dt\,,\nonumber\\
&  S_{\mathrm{non-ph}}\left[  \mathbf{\vartheta}\right]  =\int\left[
\mathcal{P}\dot{Q}+U_{p}\dot{U}_{x}-H_{\mathrm{non-ph}}\left(
\mathbf{\vartheta}\right)  \right]  dt\,, \label{Fcb.7}%
\end{align}
where%
\begin{align}
&  H_{\mathrm{non-ph}}\left(  \mathbf{\vartheta}\right)  =\lambda
_{\mathcal{P}}\mathcal{P}^{(1)}+\lambda_{U}U^{(1)}+(Q^{\left(  1\right)
}A+Q^{\left(  2...\right)  }B+\omega C)\mathcal{P}^{\left(  2...\right)
}\nonumber\\
&  +\mathcal{P}^{(2...)}D\mathcal{P}^{(2...)}+\mathcal{P}^{(2...)}%
EU^{(2...)}+U^{(2...)}GU^{(2...)}+O\left(  \vartheta^{3}\right)  \,.
\label{Fcc.1}%
\end{align}
and $A$, $B$, $C,E$ and $G$ are some matrices (see \cite{GitTy90}). We recall
that the variables $\omega$ are canonical pairs, $\omega=\left(  \omega
_{x},\omega_{p}\right)  $, of physical variables; the variables $Q$ are
nonphysical coordinates; and the variables $\Omega$ define the constraint
surface by the equations\ $\Omega=0$. At the same time, the variables $\Omega$
are divided into two groups: $\Omega=(\mathcal{P},U)$, where $U$ stands for
the complete set of SCC (they consist of canonical pairs $U=(U_{x},U_{p})$)
and $\mathcal{P}$ denotes the complete set of FCC. The momenta $\mathcal{P}$
are conjugate to the coordinates $Q.$ The special variables can be chosen so
that $\Omega=\left(  \Omega^{(1)},\Omega^{(2...)}\right)  ,$ where
$\Omega^{(1)}$ are primary and$\emph{\ }\Omega^{(2...)}$ are secondary
constraints. Respectively, $\Omega^{(1)}=(\mathcal{P}^{(1)},U^{(1)})$,
$\Omega^{(2...)}=(\mathcal{P}^{(2...)},U^{(2...)})$; $\mathcal{P=}%
(\mathcal{P}^{(1)},\mathcal{P}^{(2...)})$, $U=(U^{(1)},U^{(2...)})$;
$\mathcal{P}^{\left(  1\right)  }$ are primary FCC; $\mathcal{P}^{\left(
2...\right)  }$ are secondary FCC; $U^{\left(  1\right)  }$ are primary SCC;
$U^{\left(  2\right)  }$ are secondary SCC.

b) One can prove that the special phase-space variables can be chosen so that
the quadratic part of the non-physical part of the total Hamiltonian takes a
simple (canonical) form (see \cite{218}):%
\begin{align}
&  H_{\mathrm{non-ph}}=\sum_{a=1}^{\aleph_{\chi}}\left(  \sum_{i=1}%
^{a-1}Q^{(i|a)}\mathcal{P}^{(i+1|a)}+\lambda_{\mathcal{P}}^{a}\mathcal{P}%
^{(1|a)}\right) \nonumber\\
&  +\lambda_{U}U^{(1)}+U^{(2...)}FU^{(2...)}+O\left(  \vartheta^{3}\right)
\,. \label{Fcc.2}%
\end{align}
Here, $\left(  Q,\mathcal{P}\right)  =\left(  Q^{\left(  i|a\right)
},\mathcal{P}^{(i|a)}\right)  \,,\;\lambda_{\mathcal{P}}=\left(
\lambda_{\mathcal{P}}^{a}\right)  \,,\ \;a=1,...,\aleph_{\chi}%
\,,\;i=1,...,a\ ,$ and $F$ is a matrix. We recall that $\aleph_{\chi}$ is the
number of stages of the refined Dirac procedure (see \cite{196} and sec. 3)
that is necessary to determine all independent FCC from the orthogonal
constraint basis. We call such special phase-space variables the superspecial
phase-space variables. In the superspecial phase-space variables, the
non-physical part of the Hamiltonian action can be written as%
\begin{align}
&  S_{\mathrm{non-ph}}=S_{\mathrm{non-ph}}^{0}+S_{\mathrm{non-ph}}%
^{int}\,,\nonumber\\
&  S_{\mathrm{non-ph}}^{0}=\int\left[  \mathcal{P}\hat{\Lambda}\mathcal{Q}%
+\sum_{i=1}^{\aleph_{\chi}}\mathcal{P}^{(i|i)}\dot{Q}^{(i|i)}+\mathcal{U}%
\hat{B}\mathcal{U}\right]  dt\,,\;S_{\mathrm{non-ph}}^{int}=O\left(
\vartheta^{3}\right)  \,, \label{Fcc.15}%
\end{align}
where $\hat{\Lambda}$ and $\hat{B}$ are first-order LO with constant
coefficients, and%
\[
\mathcal{Q}=(\lambda_{\mathcal{P}}^{a},\,Q^{(i|a)}%
\,,\;\,i=1,...,a-1\,,\;a=1,...,\aleph_{\chi})\,,\;\mathcal{U}=(\lambda
_{U},U)\,.
\]
It is important to stress that $[\mathcal{Q]}=\left[  \mathcal{P}\right]  \,,$
due to the fact that $\left[  \lambda_{\mathcal{P}}\right]  =\left[
\mathcal{P}^{(1)}\right]  .$ One can see that there exist LO $\hat{\Lambda
}^{-1}$ and $\hat{B}^{-1}$ such that%
\[
\hat{\Lambda}\hat{\Lambda}^{-1}=\hat{\Lambda}^{-1}\hat{\Lambda}=1,\;\hat
{B}\hat{B}^{-1}=\hat{B}^{-1}\hat{B}=1\,.
\]

Proceeding to the proof of the statement, we consider the Hamiltonian action
$\tilde{S}_{\mathrm{H}}$ (\ref{Fcc.15}) in the superspecial phase-space
variables. The equations of motion that follow from this action have the form%
\begin{align}
\frac{\delta\tilde{S}_{\mathrm{H}}}{\delta\mathcal{U}}  &  =0\Longrightarrow
\mathcal{U}=-\hat{B}^{-1}\frac{\delta S_{\mathrm{non-ph}}^{int}}%
{\delta\mathcal{U}}\,,\;S_{\mathrm{non-ph}}^{int}=O\left(  \vartheta
^{3}\right)  \,,\nonumber\\
\;\frac{\delta\tilde{S}_{\mathrm{H}}}{\delta\mathcal{Q}}  &  =0\Longrightarrow
\mathcal{P}=-\left(  \hat{\Lambda}^{T}\right)  ^{-1}\frac{\delta
S_{\mathrm{non-ph}}^{int}}{\delta\mathcal{Q}}\,,\nonumber\\
\frac{\delta\tilde{S}_{\mathrm{H}}}{\delta\mathcal{P}}  &  =0\Longrightarrow
\mathcal{Q}=-\hat{\Lambda}^{-1}\left(  \frac{\delta S_{\mathrm{non-ph}}^{int}%
}{\delta\mathcal{P}}+T\right)  \,,\;T=\left(  T^{\left(  i|u\right)  }%
=\delta_{iu}\dot{Q}^{\left(  u|u\right)  }\right)  \,. \label{sdc.3d}%
\end{align}
These equations allow one to express all the variables $\mathcal{U}%
,\mathcal{P}$, and$\;\mathcal{Q}$ as some LF of $\omega$ and $Q^{\left(
i|i\right)  }$, at least perturbatively. Note that, in any case, the exact
solution of equations (\ref{sdc.3d}) has the form $\mathcal{U}=\mathcal{P}%
=0$,\ $\mathcal{Q}=\psi\left(  \omega,Q^{\left(  i|i\right)  }\right)  ,$
where $\psi$ is an LF of the indicated arguments. Therefore, the variables
$\mathcal{U}$, $\mathcal{P}$, and$\;\mathcal{Q}$ are auxiliary ones; see
\cite{Superflu,195}. Excluding these variables from the action $\tilde
{S}_{\mathrm{H}},$ we obtain a dynamically equivalent action $\bar
{S}_{\mathrm{H}}\left[  \omega,Q^{\left(  i|i\right)  }\right]  $. Taking into
account the fact that $\mathcal{U}=\mathcal{P}=0\Longrightarrow\Omega=0,$ and
relation (\ref{2.9}), we obtain%
\begin{equation}
\bar{S}_{\mathrm{H}}\left[  \omega,Q^{\left(  i|i\right)  }\right]  =\left.
\tilde{S}_{\mathrm{H}}\right|  _{\mathcal{U}=\mathcal{P}=0,\mathcal{Q}=\psi
}=S_{\mathrm{ph}}\left[  \omega\right]  \,. \label{sdc.3c}%
\end{equation}
One ought to keep in mind that equality (\ref{sdc.3c}) does not imply the
dynamical equivalence of the actions $\bar{S}_{\mathrm{H}}$ and
$S_{\mathrm{ph}}$ (and, therefore, equivalence of $\tilde{S}_{\mathrm{H}}$ and
$S_{\mathrm{ph}}$); see \cite{GitTy05b}.

Let a transformation $\delta\vartheta,\delta\lambda$, vanishing on the
extremals of $\tilde{S}_{\mathrm{H}}$, be a symmetry of the action $\tilde
{S}_{\mathrm{H}}$ . Consider the reduced transformation $\bar{\delta}%
\omega,\bar{\delta}Q^{\left(  i|i\right)  },$%
\[
(\bar{\delta}\omega,\bar{\delta}Q^{\left(  i|i\right)  })=\left.
(\delta\vartheta,\delta\lambda)\right|  _{\mathcal{U}=\mathcal{P}%
=0,\mathcal{Q}=\psi}\,.
\]
Obviously, the reduced transformation vanishes on the extremals of the reduced
action $\bar{S}_{\mathrm{H}}$\ and is a symmetry transformation of the action
$\bar{S}_{\mathrm{H}}$. This implies%
\[
\bar{\delta}\omega=\hat{m}\frac{\delta S_{\mathrm{ph}}}{\delta\omega}%
\,,\;\bar{\delta}Q^{\left(  i|i\right)  }=\left(  \hat{n}\frac{\delta
S_{\mathrm{ph}}}{\delta\omega}\right)  ^{\left(  i|i\right)  }\,,
\]
where $\hat{m}$ and $\hat{n}$ are some LO. The transformation $\bar{\delta
}\omega$ is obviously a symmetry transformation of the nonsingular action
$S_{\mathrm{ph}}$ that vanishes on its extremals. Therefore, according to the
assertion that was proved in the previous subsection, $\hat{m}$ is an
antisymmetric LO. Thus, the complete transformation $\bar{\delta}\omega
,\bar{\delta}Q^{\left(  i|i\right)  }$ can be represented in the form
\[
\left(
\begin{array}
[c]{l}%
\bar{\delta}\omega\\
\bar{\delta}Q^{\left(  |\right)  }%
\end{array}
\right)  =\hat{M}\left(
\begin{array}
[c]{l}%
\frac{\delta\bar{S}_{\mathrm{H}}}{\delta\omega}\\
\frac{\delta\bar{S}_{\mathrm{H}}}{\delta Q^{\left(  |\right)  }}%
\end{array}
\right)  \,,\;\hat{M}=\left(
\begin{array}
[c]{cc}%
\hat{m} & -\hat{n}^{T}\\
\hat{n} & 0
\end{array}
\right)  \,.
\]
Here, obviously, $\hat{M}$ is an antisymmetric matrix.

Finally, the transformation $\bar{\delta}\omega,\bar{\delta}Q^{\left(
i|i\right)  }$ is a trivial symmetry of the action $\bar{S}_{\mathrm{H}}.$
This implies that the extended transformation $\delta\mathbf{\vartheta}$ is a
trivial symmetry of the extended action $\tilde{S}_{\mathrm{H}}$, according to
the general statement presented in \cite{GitTy05b}.

\section{Physical functions in gauge theories}

Despite a functional arbitrariness in solutions of the equations of motion for
gauge theories, such theories can be used to describe physics. To ensure the
independence of physical quantities from the arbitrariness inherent in
solutions of a gauge theory, one imposes restrictions on a possible form of
physical functions that describe physical quantities.

It was demonstrated (see \cite{GitTy90}) that physical LF
$A_{\text{\textrm{ph}}}\left(  \mathbf{\eta}^{[]}\right)  $ in the Hamiltonian
formulation (we recall that $\mathbf{\eta=}\left(  \eta,\lambda\right)
,\;\eta=\left(  q,p\right)  $) have the following structure:%
\begin{equation}
A_{\text{\textrm{ph}}}\left(  \mathbf{\eta}^{[]}\right)  =a\left(
\eta\right)  +O\left(  \frac{\delta S_{\mathrm{H}}}{\delta\mathbf{\eta}%
}\right)  \,, \label{pia.6}%
\end{equation}
where $\chi$ stands for the complete set of FCC in the theory, and $a\left(
\eta\right)  $ is a function of the phase-space variables that obeys the
following condition:%
\begin{equation}
\{a,\chi\}=O\left(  \Phi\right)  \,. \label{pia.6a}%
\end{equation}
We are going to call conditions (\ref{pia.6}) the physicality conditions in
the Hamiltonian sense. It is precisely in this sense that one must understand
the usual statement that physical functions must commute with FCC on
extremals. In fact, these conditions of physicality are those which are
usually called the Dirac conjecture.

On the other hand, it is known that physical functions must be gauge-invariant
on the extremals; see, e.g., \cite{GitTy90}. Let $\delta_{\nu}\mathbf{\eta}$
be a gauge symmetry of the Hamiltonian action. Then, the gauge variations of
the LF $A_{\text{\textrm{ph}}}\left(  \mathbf{\eta}^{[]}\right)  $ must be
proportional to the extremals:%
\begin{equation}
\delta_{\nu}A_{\text{\textrm{ph}}}\left(  \mathbf{\eta}^{[]}\right)  =O\left(
\frac{\delta S_{\mathrm{H}}}{\delta\mathbf{\eta}}\right)  \,. \label{pia.7}%
\end{equation}
Such a condition will be called the physicality condition in a gauge sense.
Until now, it has not been strictly proved whether the two definitions
(\ref{pia.6}), (\ref{pia.6a}) and (\ref{pia.7}) are equivalent. Using the
structure of arbitrary gauge transformation established in Section 4, we are
going to prove below an equivalence of these two definitions.

Let an LF $A_{\text{\textrm{ph}}}\left(  \mathbf{\eta}^{[]}\right)  $ be
physical in the Hamiltonian sense, i.e., it obeys (\ref{pia.6}) and
(\ref{pia.6a}). Consider its gauge variation $\delta_{\nu}A_{\text{\textrm{ph}%
}}$. Such a variation has the form%
\begin{equation}
\delta_{\nu}A_{\text{\textrm{ph}}}=\delta_{\nu}a\left(  \eta\right)  +O\left(
\frac{\delta S_{\mathrm{H}}}{\delta\mathbf{\eta}}\right)  \,. \label{pia.15}%
\end{equation}
Here, we have used the fact that gauge variations of extremals are
proportional to extremals. Taking into account (\ref{3.5}) and (\ref{3.15}),
we can see that the variation $\delta_{\nu}a_{\text{\textrm{ph}}}$ is
proportional to extremals:%
\[
\delta a=\{a,G_{\nu}\}=O\left(  \left\{  a,\chi\right\}  \right)  +O\left(
\frac{\delta S_{\mathrm{H}}}{\delta\mathbf{\eta}}\right)  \,.
\]
Then condition (\ref{pia.6a}) implies (\ref{pia.7}).

Let an LF $A_{\text{\textrm{ph}}}\left(  \mathbf{\eta}^{[]}\right)  $ be now
physical in the gauge sense, i.e., it obeys (\ref{pia.7}). We can always
represent any LF of $\mathbf{\eta}^{[]}$ in terms of a function of the
variables $\eta$ and $\lambda_{\chi}^{[]}=\left(  \lambda^{a[]}\right)  $ and
a function proportional to extremals of the type $J$ (see (\ref{2.10})). Thus,
one can always write
\begin{equation}
A_{\text{\textrm{ph}}}\left(  \mathbf{\eta}^{[]}\right)  =f(\eta,\lambda
_{\chi}^{[]})+O(J)\,. \label{pia.16}%
\end{equation}
The condition (\ref{pia.7}), with allowance made for (\ref{3.5}) and
(\ref{3.15}), implies the equation%
\begin{equation}
\{f,G_{\nu}\}+\sum_{m=0}^{m_{\max}}\frac{\partial f}{\partial\lambda^{a[m]}%
}\delta_{\nu}\lambda^{a[m]}=O\left(  \frac{\delta S_{\mathrm{H}}}%
{\delta\mathbf{\eta}}\right)  \label{sa}%
\end{equation}
for the function $f$. We recall that the variations $\delta_{\nu}\lambda^{a}$
are given by expression (\ref{3.19}). Let us consider such terms in the LHS of
(\ref{sa}) that contain (proportional to) the highest time derivatives
$\nu_{(a)}^{[a+m_{\max}]}$ of the gauge parameters $\nu_{(a)}$. Since the
gauge charge does not contain such derivatives (see (\ref{3.15})), these terms
have the form%
\begin{equation}
\sum_{a,b=1}^{\aleph_{\chi}}\frac{\partial f}{\partial\lambda^{a[m_{\max}]}%
}\mathcal{D}^{ab}\nu_{(b)}^{[b+m_{\max}]}=O\left(  \frac{\delta S_{\mathrm{H}%
}}{\delta\mathbf{\eta}}\right)  \,, \label{pia.17}%
\end{equation}
and are proportional to extremals due to (\ref{sa}). In turn, (\ref{pia.17})
implies the relation%
\[
\frac{\partial f}{\partial\lambda^{a[m_{\max}]}}=O\left(  \frac{\delta
S_{\mathrm{H}}}{\delta\mathbf{\eta}}\right)
\]
due to the nonsingularity of the matrix $\mathcal{D}$. Similarly, we can
verify that on extremals, the function $f$ does not depend on the variables
$\lambda_{\chi}^{[]}$, i.e.,%
\begin{equation}
f(\eta,\lambda_{\chi}^{[]})=a\left(  \eta\right)  +O\left(  \frac{\delta
S_{\mathrm{H}}}{\delta\mathbf{\eta}}\right)  \,. \label{pia.18}%
\end{equation}
Therefore, a physical (in a gauge sense) LF $A_{\text{\textrm{ph}}}\left(
\mathbf{\eta}^{[]}\right)  $ must have the form (\ref{pia.6}). Considering
equation (\ref{pia.7}) for such a function, and taking into account
(\ref{3.19}), we obtain%
\begin{equation}
\,\{a,G_{\nu}\}=O\left(  \frac{\delta S_{\mathrm{H}}}{\delta\mathbf{\eta}%
}\right)  \,, \label{pia.19}%
\end{equation}
which implies%
\begin{equation}
\left(  \sum_{k=1}^{\aleph_{\chi}}\sum_{a=k}^{\aleph_{\chi}}\right)  \left(
\sum_{m=1}^{\aleph_{\chi}}\sum_{b=m}^{\aleph_{\chi}}\right)  \left\{
a,\chi^{\left(  k|a\right)  }\right\}  \mathcal{C}_{ka}^{mb}\nu_{(b)}%
^{[m-1]}=O\left(  \frac{\delta S_{\mathrm{H}}}{\delta\mathbf{\eta}}\right)
\,. \label{pia.19a}%
\end{equation}
Taking into account the facts that the matrix $\mathcal{C}$ is invertible, the
gauge parameters $\nu$ are arbitrary functions of time, and thus all
$\nu^{\lbrack]}$ are\ independent, one can derive from relation (\ref{pia.19}%
)
\begin{equation}
\left\{  a,\chi\right\}  =O\left(  \frac{\delta S_{\mathrm{H}}}{\delta
\mathbf{\eta}}\right)  \Longrightarrow\left\{  a,\chi\right\}  =O(\Phi)\,.
\label{pia.20}%
\end{equation}
Relations (\ref{pia.16}), (\ref{pia.18}) and (\ref{pia.20}) imply that an LF
$A_{\text{\textrm{ph}}}\left(  \mathbf{\eta}^{[]}\right)  $ which is physical
in the gauge sense is physical in the Hamiltonian sense as well. Thus, the
equivalence of the two definitions of physicality condition is proved.

\section{Examples}

I. As an example of how gauge symmetries can be recovered from the constraint
structure in the Hamiltonian formulation, we consider a field model which
includes a set of Yang--Mills vector fields $\mathcal{A}_{\mu}^{a}$ ,
$a=1,...,r,$ and a set of spinor fields $\psi^{\alpha}=\left(  \psi
_{i}^{\alpha},\;i=1,...,4\right)  ,$%
\begin{align}
&  S=\int\mathcal{L}dx\,,\;\mathcal{L}=-\frac{1}{4}G_{\mu\nu}^{a}G^{\mu\nu
a}+i\bar{\psi}^{\alpha}\gamma^{\mu}\nabla_{\mu\beta}^{\alpha}\psi^{\beta
}-V(\psi,\bar{\psi})\,,\nonumber\\
&  G_{\mu\nu}^{a}=\partial_{\mu}\mathcal{A}_{\nu}^{a}-\partial_{\nu
}\mathcal{A}_{\mu}^{a}+f_{bc}^{a}\mathcal{A}_{\mu}^{b}\mathcal{A}_{\nu}%
^{c}\,,\;\;\nabla_{\mu\beta}^{\alpha}=\partial_{\mu}\delta_{\beta}^{\alpha
}-iT_{a\beta}^{\alpha}\mathcal{A}_{\mu}^{a}\,, \label{ex.1}%
\end{align}
where $V$ is a local polynomial in the field, which contains no derivatives.
The model is based on a certain global Lie group $G,$%
\begin{align*}
&  \psi\left(  x\right)  \overset{g}{\rightarrow}\exp\left(  i\nu^{\alpha
}T_{a}\right)  \psi\left(  x\right)  ,\;g\in G\,,\;\nu^{a},\;a=1,...,r,\\
&  T_{a}=T_{a}^{+}\,,\;\left[  T_{\alpha},T_{b}\right]  =if_{ab}^{c}%
T_{c}\,,\;\;f_{ab}^{k}f_{kc}^{n}+f_{bc}^{k}f_{ka}^{n}+f_{ca}^{k}f_{kb}%
^{n}=0\,.
\end{align*}

For $V=0,$ the action is invariant under the gauge transformations ($\nu
^{a}=\nu^{a}\left(  x\right)  $)%
\begin{equation}
\delta\mathcal{A}_{\mu}^{a}=D_{\mu b}^{a}\nu^{b}\,,\;\delta\psi=iT_{a}\psi
\nu^{a}\,,\;D_{\mu b}^{a}=\partial_{\mu}\delta_{b}^{a}+f_{cb}^{a}%
\mathcal{A}_{\mu}^{c}\,. \label{ex.2}%
\end{equation}
We assume the polynomial $V$ to be such that the entire action (\ref{ex.1}) is
invariant also under the transformations (\ref{ex.2}). Below, we relate the
symmetry structure of the model to its constraint structure. To this end, we
first reveal the constraint structure.

Proceeding to the Hamiltonian formulation, we introduce the momenta%
\[
p_{0a}=\frac{\partial\mathcal{L}}{\partial\mathcal{\dot{A}}^{0a}}%
=0\,,\;p_{ia}=\frac{\partial\mathcal{L}}{\partial\mathcal{\dot{A}}^{ia}%
}=G_{i0}^{a}\,,\;p_{\psi}=\frac{\partial_{r}\mathcal{L}}{\partial\dot{\psi}%
}=i\bar{\psi}\gamma^{0}\,,\;p_{\bar{\psi}}=\frac{\partial_{r}\mathcal{L}%
}{\partial\overset{\cdot}{\bar{\psi}}}=0\,.
\]
Hence, there exists a set of primary constraints$\;\Phi^{(1)}=\left(  \chi
_{a}^{(1)},\varphi_{\sigma}^{(1)},\sigma=1,2\right)  =0,$ where%
\[
\chi_{a}^{(1)}=p_{0a}\,,\;\varphi_{1}^{(1)}=p_{\psi}-i\bar{\psi}\gamma
^{0}\,,\;\varphi_{2}^{(1)}=p_{\bar{\psi}}\,.
\]
The total Hamiltonian reads $H^{\left(  1\right)  }=\int\mathcal{H}%
^{(1)}d\mathbf{x}$\textbf{,}%
\[
\mathcal{H}^{(1)}=\frac{1}{2}p_{ia}^{2}+\frac{1}{4}G_{ik}^{a2}-p_{\psi}%
\gamma^{0}\gamma^{k}\nabla_{k}\psi+\mathcal{A}^{0a}(D_{ib}^{a}p_{ib}-\bar
{\psi}\gamma^{0}T_{a}\psi)+V+\lambda_{\chi}^{a}\chi_{a}^{(1)}+\lambda
_{\varphi}^{\sigma}\varphi_{\sigma}^{(1)}\,.
\]
By performing the Dirac procedure, one can verify that there appear only
secondary constraints $\chi_{a}^{(2)}=0,$
\begin{align*}
&  \left\{  \varphi_{\sigma}^{(1)},H^{\left(  1\right)  }\right\}
=0\Longrightarrow\lambda_{\varphi}^{\sigma}=\bar{\lambda}_{\varphi}^{\sigma
}\left(  \mathcal{A},\psi,\bar{\psi}\right)  \,,\\
&  \left\{  \chi_{a}^{(1)},H^{\left(  1\right)  }\right\}  =0\Longrightarrow
\chi_{a}^{(2)}=D_{ib}^{a}p_{ib}+i\left(  p_{\psi}T_{a}\psi+p_{\bar{\psi}}%
\bar{T}_{a}\bar{\psi}\right)  \,,\;\left(  \bar{T}_{a}\right)  _{\beta
}^{\alpha}=-\gamma^{0}\left(  T_{a}^{\ast}\right)  _{\beta}^{\alpha}\gamma
^{0}\,.
\end{align*}
All constraints $\varphi$ are of second-class, and all constraints $\chi$ are
of first-class. It turns out that the complete set of constraints already
forms an orthogonal constraint basis, namely,%
\[
\varphi^{(1|1)}\equiv\varphi^{(1)},\;\chi^{(1|2)}\equiv\chi^{(1)}%
\,,\;\chi^{(2|2)}\equiv\chi^{(2)},
\]
and there are no constraints $\chi^{(1|1)},$%
\[%
\begin{array}
[c]{ccccc}%
\varphi^{(1|1)} & \rightarrow & \bar{\lambda} &  & \\
\chi^{(1|2)} & \rightarrow & \chi^{(2|2)} & \rightarrow & O\left(
\Phi\right)
\end{array}
.
\]
According to the general considerations, we chose the gauge charge in the
form
\[
G=\int\left[  \nu^{a}\chi_{a}^{(2|2)}+C^{a}\chi_{a}^{(1|2)}\right]
d\mathbf{x}\,,\;C^{a}=\left(  c_{b}^{a}\nu^{b}+d_{b}^{a}\dot{\nu}^{b}\right)
\,.
\]
Solving the symmetry equation, we obtain $C^{a}=\dot{\nu}^{a}-\nu
^{c}\mathcal{A}^{0b}f_{cb}^{a}=D_{0b}^{a}\nu^{b}\,.$ Thus,
\begin{align*}
&  G=\int\left[  p_{\mu a}D_{b}^{\mu a}\nu^{b}+i\left(  p_{\psi}T_{a}%
\psi+p_{\bar{\psi}}\bar{T}_{a}\bar{\psi}\right)  \nu^{a}\right]
d\mathbf{x\,,}\\
&  \delta\mathcal{A}_{\mu}^{a}=\left\{  \mathcal{A}_{\mu}^{a},G\right\}
=D_{\mu b}^{a}\nu^{b}\,,\;\delta\psi=\left\{  \psi,G\right\}  =iT_{a}\psi
\nu^{a}\,.\,
\end{align*}

II. Below, we represent a gauge model in which the gauge charge must be
constructed with the help of both FCC and SCC.

Consider a Hamiltonian action $S_{\mathrm{H}}$ that depends on phase-space
variables $q_{i},p_{i},\;i=1,2,$ and $x_{a},\pi_{a},\;\alpha=1,2,$ and on two
Lagrange multipliers $\lambda_{\pi}$ and $\lambda_{p}\,,$%
\begin{align}
&  S_{\mathrm{H}}=\int\left[  p_{i}\dot{q}_{i}+\pi_{\alpha}\dot{x}_{\alpha
}-H^{(1)}\right]  dt\,,\;H^{(1)}=H_{0}^{(1)}+V\,,\nonumber\\
&  H_{0}^{(1)}=\frac{1}{2}\pi_{2}^{2}+x_{1}\pi_{2}+\frac{1}{2}p_{2}%
^{2}+\frac{1}{2}q_{2}^{2}+q_{1}p_{2}+\lambda_{\pi}\pi_{1}+\lambda_{p}%
p_{1}\,,\;\;V=x_{1}q_{2}^{2}\,. \label{spr.1}%
\end{align}
In what follows, we denote all the variables by $\mathbf{\eta}=\left(
x,\pi,q,p,\lambda\right)  .$ The model has two primary constraints $\pi_{1}$
and $p_{1}.$ One can consider $V$ as the interaction Hamiltonian. It is easy
to verify that a complete set of constraints can be chosen as $\chi=\left(
\pi_{1},\pi_{2}\right)  $ and $\varphi=\left(  q_{1},q_{2},p_{1},p_{2}\right)
.$ Here, $\chi$ are FCC and $\varphi$ are SCC. As was already mentioned, gauge
symmetries of the action $S_{\mathrm{H}}$ have gauge charges that must be
constructed with the help of both FCC and SCC. To demonstrate this fact, let
us try to solve the symmetry equation with the gauge charge that is
proportional only to FCC. The general form of such a charge can be written as%
\begin{equation}
G=A^{a}\tilde{\chi}_{a}\,. \label{spr.2}%
\end{equation}
Here, $\tilde{\chi}_{a}\,,\;a=1,2$, are FCC,%
\begin{equation}
\tilde{\chi}_{a}=\pi_{a}+\psi_{a}+O(\eta^{3})\,,\;\;\psi_{a}=\psi
_{a}(q,p)=O(q^{2},qp,p^{2})\,, \label{spr.3}%
\end{equation}
and $A^{a}$ are some functions of the form%
\begin{equation}
A^{a}=A_{0}^{a}+A_{1}^{a}(\mathbf{\eta}^{[]})+O(\mathbf{\eta}^{2})\,,
\label{spr.4}%
\end{equation}
where $A_{0}^{a}$ are some functions of time only,\footnote{We recall that all
LF under consideration may depend on time explicitly; however, we do not
include $t$ in the arguments.} and $A_{1}^{a}$ are certain linear LF of the
indicated arguments. Thus,%
\[
G=G_{0}+G_{1}+O(\mathbf{\eta}^{3})\,,\;\;G_{0}=A_{0}^{a}\pi_{a}\,,\;\;G_{1}%
=A_{0}^{a}\psi_{a}+A_{1}^{a}\pi_{a}\,.
\]
The symmetry equation in the case under consideration reads
\begin{equation}
\hat{\partial}_{t}G+\{G,H^{(1)}\}=O\left(  \Phi^{\left(  1\right)  }\right)
\,,\;\;\hat{\partial}_{t}=\partial_{t}+\lambda^{\left[  m+1\right]
}\frac{\partial}{\partial\lambda^{\left[  m\right]  }}\,. \label{spr.5}%
\end{equation}
Considering this equation in the zero and first order with respect to the
interaction $V$, we obtain%
\begin{align}
&  \hat{\partial}_{t}G_{0}+\{G_{0},H_{0}^{(1)}\}=O\left(  \Phi^{\left(
1\right)  }\right)  \,,\label{spr.7}\\
&  \hat{\partial}_{t}G_{1}{+}\{G_{0},V\}+\{G_{1},H_{0}^{(1)}\}=O\left(
\Phi^{\left(  1\right)  }\right)  \,. \label{spr.6}%
\end{align}
Equation (\ref{spr.7}) implies
\begin{equation}
\pi_{1}\hat{\partial}_{t}A_{0}^{1}+\pi_{2}\hat{\partial}_{t}A_{0}^{2}-\pi
_{2}A_{0}^{1}=O(\pi_{1})\Longrightarrow A_{0}^{1}=\hat{\partial}_{t}A_{0}%
^{2}\,. \label{spr.8}%
\end{equation}
We can see that $A_{0}^{2}$ enter in the solution of the symmetry equation as
an arbitrary function of time. In fact, we can identify this function with a
time dependent gauge parameter. Taken on the constraint surface $\pi_{a}=0,$
equation (\ref{spr.6}) reads
\begin{equation}
\psi_{1}\hat{\partial}_{t}^{2}A_{0}^{2}+\left(  \hat{\partial}_{t}\psi
_{1}+\psi_{2}-q_{2}^{2}+\{\psi_{1},H_{0}^{(1)}\}\right)  \hat{\partial}%
_{t}A_{0}^{2}+\left(  \hat{\partial}_{t}\psi_{2}+\{\psi_{2},H_{0}%
^{(1)}\}\right)  A_{0}^{2}=O(p_{1})\,. \label{spr.9}%
\end{equation}
Since the constraints $\psi$ and the Hamiltonian $H_{0}^{(1)}$ do not depend
on the gauge parameter, we obtain from (\ref{spr.9}) and (\ref{spr.3})%
\[
\psi_{1}=O(p_{1})=(\alpha_{1}q_{1}+\alpha_{2}q_{2}+\alpha_{3}p_{1}+\alpha
_{4}p_{2})p_{1}\,,
\]
where $\alpha$ are some functions of time. With allowance for this result, we
obtain from (\ref{spr.9}) and (\ref{spr.3})
\begin{equation}
\psi_{2}=q_{2}^{2}+(\alpha_{1}q_{1}+\alpha_{2}q_{2}+\alpha_{4}p_{2}%
)p_{2}+(\beta_{1}q_{1}+\beta_{2}q_{2}+\beta_{3}p_{1}+\beta_{4}p_{2})p_{1}\,,
\label{spr.10}%
\end{equation}
where $\beta$ are some functions of time. Similarly, we have
\begin{equation}
\hat{\partial}_{t}\psi_{2}+\{\psi_{2},H_{0}^{(1)}\}=(2-\alpha_{1})q_{1}%
q_{2}+\alpha_{1}\lambda_{p}p_{2}+\Delta=O(p_{1})\,, \label{spr.11}%
\end{equation}
where $\Delta$ does not contain terms of the form $q_{1}q_{2}$ and
$\lambda_{p}p_{2}.$ Therefore, relation (\ref{spr.11}) is contradictory.

Thus, we have demonstrated that the symmetry equation has no solutions for the
gauge charge of the form (\ref{spr.2}). The gauge charge must depend on all
constraints, both FCC and SCC. An example of such a charge (solution of the
symmetry equation (\ref{spr.5})) is%
\begin{equation}
G=(\pi_{2}+q_{2}^{2}+2q_{1}p_{2}+2q_{2}p_{1}+2\lambda_{p}p_{1})\nu\left(
t\right)  +(\pi_{1}+2q_{1}p_{1})\dot{\nu}\left(  t\right)  +O(\mathbf{\eta
}^{3})\,, \label{spr.12}%
\end{equation}
where $\nu\left(  t\right)  $ is the gauge parameter.

III. As another example, we present below a model with FCC. Here, any
nontrivial gauge symmetry of the Hamiltonian action $S_{\mathrm{H}}$ has a
gauge charge which depends on Lagrange multipliers. The Hamiltonian
$S_{\mathrm{H}}$ and Lagrangian action $S$ of the model have the form
\begin{align}
&  S_{\mathrm{H}}=\int dt\left(  p_{x^{i}}\dot{x}^{i}+p_{y^{i}}\dot{y}%
^{i}+p_{z^{i}}\dot{z}^{i}-H^{(1)}\right)  \,,\nonumber\\
&  S=\frac{1}{2}\int dt\left[  (\dot{y}^{i}+x^{i})^{2}+(\dot{z}^{i}%
+y^{i}+g\epsilon_{jk}^{i}x^{j}y^{k})^{2}\right]  \,, \label{spr.12a}%
\end{align}
where%
\begin{align*}
&  H^{(1)}=H+\lambda^{i}\Phi_{i}^{(1)}\,,\;H=\frac{1}{2}\left(  p_{y^{i}}%
^{2}+p_{z^{i}}^{2}\right)  -x^{i}p_{y^{i}}-y^{i}p_{z^{i}}-g\epsilon_{jk}%
^{i}x^{j}y^{k}p_{z^{i}}\,,\\
&  \Phi_{i}^{(1)}=p_{x^{i}}\,,\;\;\epsilon_{jk}^{i}=-\epsilon_{kj}%
^{i}\,,\;\epsilon_{12}^{i}=1,\;\forall i\;;\;i,j,k=1,2\,,
\end{align*}
and $g$ is a constant.

Considering the consistency conditions for the primary constraints $\Phi
_{i}^{(1)}$, we find second-stage constraints $\Phi_{i}^{(2)},$%
\[
\{\Phi_{i}^{(1)},H\}=p_{y^{i}}+g\epsilon_{il}^{j}y^{l}p_{z^{j}}%
=0\Longrightarrow\Phi_{i}^{(2)}=p_{y^{i}}+g\epsilon_{il}^{j}y^{l}p_{z^{j}}\,.
\]
Similarly, we find third-stage constraints $\Phi_{i}^{(3)}$,
\[
\Phi_{i}^{(3)}=\{\Phi^{(2)},H\}=\left(  \delta_{i}^{j}+2g\epsilon_{li}%
^{j}x^{l}+g\epsilon_{il}^{j}p_{y^{l}}\right)  p_{z^{j}}\,.
\]
No more constraints appear from the consistency conditions. All the
constraints are FCC. We can replace the constraints $\Phi_{i}^{(1)}$,
$\Phi_{i}^{(2)},$ and $\Phi_{i}^{(3)}$ by an equivalent set of FCC,
$T_{i}^{(1)},T_{i}^{(2)},$ and $T_{i}^{(3)},$ which is%
\[
T_{i}^{(1)}=p_{x^{i}}\,,\;T_{i}^{(2)}=p_{y^{i}}\,,\;T_{i}^{(3)}=p_{z^{i}}\,.
\]

Let us suppose that a $\delta\eta=\{\eta,G\}$ and $\delta\lambda$ is a gauge
transformation of the action $S_{\mathrm{H}}$, with the gauge charge $G$ that
does not depend on $\lambda.$ We can present such a charge as
\begin{equation}
G=\sum_{l=0}^{N}u^{\left[  l\right]  }(t)G_{l}(\eta)\,.\label{spr.12b}%
\end{equation}
where $u\left(  t\right)  $ is a function of time. The symmetry equation, in
the case under consideration, has the form%
\[
\partial_{t}G+\{G,H\}+\{G,p_{x^{i}}\}\lambda=O(p_{x})\,.
\]
Since the gauge charge does not depend on $\lambda$, for $\lambda=0,$ we
obtain from this equation%
\begin{equation}
\{G,H\}+\partial_{t}G=O(p_{x})\,.\label{402}%
\end{equation}
Then%
\begin{equation}
\{G,p_{x^{i}}\}\lambda=O(p_{x})\Longrightarrow\{G,p_{x^{i}}\}=O(p_{x}%
)\,.\label{403}%
\end{equation}

It follows from (\ref{402}) that the function $G_{N}(\eta)$\thinspace can be
represented in the form
\[
G_{N}=A^{i}p_{x^{i}}+O(\Phi^{2})\,,
\]
where $A^{i}$ are functions of the coordinates only. Then, considering the
terms proportional to $u^{[N]}$ in the symmetry equation, we can see that the
function $G_{N-1}(\eta)$\thinspace can be presented in the form
\[
G_{N-1}=-A^{i}\Phi_{i}^{(2)}+B^{i}p_{x^{i}}+O(\Phi^{2})\,,
\]
where $B^{i}$ are functions of the coordinates only. Relation (\ref{403})
implies $\partial_{x^{j}}A^{i}=0.$ Considering the terms proportional to
$u^{[N-1]}$ in the symmetry equation, we can see that the function
$G_{N-2}(\eta)$\thinspace can be presented in the form
\[
G_{N-2}=-D^{i}\Phi_{i}^{(2)}+A^{i}(\delta_{i}^{j}+2g\epsilon_{li}^{j}%
x^{l})p_{z^{j}}+C^{i}p_{x^{i}}+O(\Phi^{2})\,,
\]
where
\[
D^{i}\equiv B^{i}-\partial_{t}A^{i}-(x^{l}\partial_{y^{l}}+y^{l}%
\partial_{z^{l}}+g\epsilon_{mn}^{l}x^{m}y^{n}\partial_{z^{l}})A^{i}\,,
\]
and $C^{i}$ are functions of the coordinates only. Considering the terms
proportional to $u^{[N-2]}$ in the symmetry equation, we obtain
\[
\epsilon_{ij}^{l}A^{i}=0\Longrightarrow A^{i}=0\Longrightarrow D^{i}%
=B^{i}\,;\;\;\partial_{x^{j}}D^{i}=0\Longrightarrow\partial_{x^{j}}B^{i}=0\,.
\]
Therefore,
\[
G_{N-2}=-B^{i}\Phi_{i}^{(2)}+C^{i}p_{x^{i}}+O(\Phi^{2})\,.
\]
Proceeding in the same manner, we obtain, at the final stage of the procedure,%
\[
G_{0}=-M^{i}\Phi_{i}^{(2)}+K^{i}p_{x^{i}}+O(\Phi^{2}),
\]
where $M^{i}$ and $K^{i}$ are functions of the coordinates only. Substituting
the functions $G_{0},...,G_{N-1},$ and $G_{N}$ in the above form back to the
symmetry equation, we obtain
\[
u(t)\left[  M^{i}\Phi_{i}^{(3)}+(K^{i}+O(M))\Phi_{i}^{(2)}\right]  +O(\Phi
^{2})=O(p_{x}).
\]
Since the constraints $\Phi^{(2)}$ and $\Phi^{(3)}$ are independent, we
conclude that $M=K=0$, which means, in turn, that the charge (\ref{spr.12b})
vanishes quadratically on the extremals, i.e., $G=O(\Phi^{2}).$ Therefore, the
symmetry under consideration is trivial, and a real gauge charge must contain
Lagrange multipliers.

IV. Finally, we present below a model with SCC, in which the canonical charge
depends on Lagrange multipliers. This model is described by a Hamiltonian
action $S_{\mathrm{H}}$ that depends on phase-space variables $q_{a}%
,p_{a},\;a=1,2,$ and $x_{a},\pi_{a},\;\alpha=1,2,3,$ as well as on a Lagrange
multiplier $\lambda\,,$%
\begin{align}
&  S_{\mathrm{H}}=\int\left[  p_{a}\dot{q}_{a}+\pi_{\alpha}\dot{x}_{\alpha
}-H^{(1)}\right]  dt\,,\;\;H^{(1)}=H_{0}^{\left(  1\right)  }+V\,,\;V=fq_{1}%
x_{1}x_{3}^{2}\,,\nonumber\\
&  H_{0}^{\left(  1\right)  }=\frac{1}{2}\left(  q_{a}^{2}+p_{a}^{2}\right)
+x_{1}\pi_{2}+x_{2}\pi_{3}+\frac{1}{2}x_{3}^{2}+\frac{1}{2}\pi_{2}%
^{2}+\frac{1}{2}\pi_{3}^{2}+\lambda\pi_{1}\,, \label{spr.13}%
\end{align}
The model has one primary constraint, $\varphi^{(1)}=\pi_{1}$. One can
consider $V$ as the interaction Hamiltonian with a coupling constant $f$. To
demonstrate that the symmetries of the action $S_{\mathrm{H}}$ have charges
that must depend on Lagrange multipliers, one has to find all the constraints
of the model and then to analyze the symmetry equation. We leave this problem
to the reader's consideration.

\section{Summary}

Below, we summarize the main results of the present article:

A constructive procedure of solving the symmetry equation with the help of a
so-called orthogonal constraint basis is proposed.

Using such a procedure, we can determine all the gauge transformations of a
given action. In particular, we represent the gauge charge as a decomposition
in constraints of the theory; see (\ref{3.2}). Thus, we establish a relation
between the constraint structure of the theory and the structure of its gauge
transformations. We stress that, in the general case, the gauge charge cannot
be constructed with the help of some complete set of FCC alone, since its
decomposition contains SCC as well. The gauge charge necessarily contains a
part that vanishes linearly in the FCC, and the remaining part of the gauge
charge vanishes quadratically on the extremals. With accuracy up to a trivial
transformation, any gauge transformation can be represented in the form
(\ref{3.5}), (\ref{3.6}), with the gauge charge (\ref{3.2}). The gauge charge
contains time derivatives of the gauge parameters whenever there exist
secondary FCC. Namely, the order of the highest time derivative that enters
the gauge charge is equal to $\aleph_{\chi}-1,$\ where $\aleph_{\chi}$\ is the
number of the last stage when new FCC still appear.

The above-mentioned procedure of solving the symmetry equation allows one to
analyze the structure of any infinitesimal Noether symmetry. Thus, one can see
that any infinitesimal Noether symmetry can be represented as a sum of three
kinds of symmetries: global, gauge, and trivial symmetries. The global part of
a symmetry does not vanish on the extremals, and the corresponding charge does
not vanish on the extremals either. The gauge part of a symmetry does not
vanish on the extremals, but the gauge charge vanishes on them. The trivial
part of any symmetry vanishes on the extremals, and the corresponding charge
vanishes quadratically on the extremals. The above division is not unique. In
particular, the determination of the global charge from the corresponding
symmetry equation, and thus the determination of the global part of a
symmetry, is ambiguous. However, the ambiguity in the global part of a
symmetry transformation is always the sum of a gauge transformation and a
trivial transformation. The reduction of symmetry variations to extremals are
global canonical symmetries of the physical action, whose conserved charge is
the reduction of the complete conserved charge to the extremals. Any global
canonical symmetry of the physical action can be extended to a nontrivial
global symmetry of the complete Hamiltonian action.

We note that in our procedure of solving the symmetry equation, the generators
of canonical global and gauge symmetries may depend on Lagrange multipliers
and their time derivatives. This happens in the case when the number of stages
in the Dirac procedure is more than two. We have presented examples of models
where generators of canonical and gauge symmetries necessarily depend on
Lagrange multipliers.

We have proved that any infinitesimal Noether symmetry that vanishes on the
extremals is a trivial symmetry.

Finally, using the revealed structure of arbitrary gauge transformation, we
have strictly proved an equivalence of two definitions of physicality
condition in gauge theories. One of them states that physical functions are
gauge-invariant on the extremals, and the other requires that physical
functions commute with FCC (the Dirac conjecture).

\begin{acknowledgement}
Gitman is grateful to the Brazilian foundations FAPESP and CNPq for permanent
support; Tyutin thanks RFBR 05-02-17217, and LSS-1578.2003.2 for partial support.
\end{acknowledgement}

\end{document}